\documentclass[a4paper,aps,prb,reprint,floatfix,superscriptaddress,amsmath,amssymb,amsfonts,longbibliography]{revtex4-1}

\usepackage[scaled=0.9]{helvet}

\usepackage[utf8]{inputenc}
\usepackage{textcomp}
\usepackage{amsbsy}
\usepackage{amstext}
\usepackage[pdftex]{graphicx}
\usepackage{esint}

\makeatletter

\usepackage{upgreek}
\usepackage[dvipsnames]{xcolor}
\usepackage{soul}
\usepackage{xspace}
\usepackage[rightcaption]{sidecap}
\usepackage{units}

\DeclareMathAlphabet{\mathcal}{OMS}{cmsy}{m}{n} 

\setcitestyle{numbers,square}

\makeatother

\begin{document}
\title{Determination of the carrier diffusion length in GaN from cathodoluminescence maps around threading dislocations: fallacies and opportunities}
\author{Vladimir M. Kaganer}
\affiliation{Paul-Drude-Institut f\"ur Festk\"orperelektronik, Leibniz-Institut
im Forschungsverbund Berlin e.\,V., Hausvogteiplatz 5--7, 10117
Berlin, Germany}
\author{Jonas L\"ahnemann}
\affiliation{Paul-Drude-Institut f\"ur Festk\"orperelektronik, Leibniz-Institut
im Forschungsverbund Berlin e.\,V., Hausvogteiplatz 5--7, 10117
Berlin, Germany}
\author{Carsten Pf\"uller}
\affiliation{Paul-Drude-Institut f\"ur Festk\"orperelektronik, Leibniz-Institut
im Forschungsverbund Berlin e.\,V., Hausvogteiplatz 5--7, 10117
Berlin, Germany}
\author{Karl K. Sabelfeld}
\affiliation{Institute of Computational Mathematics and Mathematical Geophysics,
Russian Academy of Sciences, Lavrentiev Prosp.\ 6, 630090 Novosibirsk,
Russia}
\author{Anastasya E. Kireeva}
\affiliation{Institute of Computational Mathematics and Mathematical Geophysics,
Russian Academy of Sciences, Lavrentiev Prosp.\ 6, 630090 Novosibirsk,
Russia}
\author{Oliver Brandt}
\affiliation{Paul-Drude-Institut f\"ur Festk\"orperelektronik, Leibniz-Institut
im Forschungsverbund Berlin e.\,V., Hausvogteiplatz 5--7, 10117
Berlin, Germany}
\date{\today}
\begin{abstract}
We investigate, both theoretically and experimentally, the drift,
diffusion, and recombination of excitons in the strain field of an
edge threading dislocation intersecting the GaN\{0001\} surface. We
calculate and measure hyperspectral cathodoluminescence maps around
the dislocation outcrop for temperatures between 10 to 200\,K. Contrary
to common belief, the cathodoluminescence intensity contrast is only
weakly affected by exciton diffusion, but is caused primarily by exciton
dissociation in the piezoelectric field at the dislocation outcrop.
Hence, the extension of the dark spots around dislocations in the
luminescence maps cannot be used to determine the exciton diffusion
length. However, the cathodoluminescence energy contrast, reflecting
the local bandgap variation in the dislocation strain field, does
sensitively depend on the exciton diffusion length and hence enables
its experimental determination.
\end{abstract}
\maketitle

\section{Introduction}

The minority carrier, ambipolar, or exciton diffusion length is the
quantity that governs all scenarios where electrons and holes or excitons
diffuse and recombine, and is as such one of the crucial parameters
that controls the behavior of semiconductor devices. A popular method
to experimentally determine the diffusion length relies on the perception
that threading dislocations in semiconductors are line defects that
act as nonradiative sinks for minority charge carriers. The zone of
reduced luminescence intensity around the dislocation is thus related
to the carrier or exciton diffusion length \citep{lax78,donolato78,donolato79,donolato85,donolato98,jakubowicz85,jakubowicz86,pasemann91}.

This method has been frequently employed to determine the exciton
diffusion length in GaN(0001) films, for which threading dislocations
are visible as dark spots in cathodoluminescence (CL) or electron-beam
induced current (EBIC) maps \citep{rosner97,shmidt02,nakaij05,pauc06,yakimov07,ino08,yakimov10,yakimov15}.
In the majority of previous experimental work, the diffusion length
was extracted from the available data in a simple phenomenological
fashion that had no sound physical foundation. We have recently derived
a rigorous solution for the intensity contrast around threading dislocations
in GaN\{0001\} considering fully three-dimensional generation, diffusion
and recombination of excitons in the presence of a surface and a dislocation
both possessing finite recombination strengths \citep{sabelfeld17CL}.
Our study has shown that the phenomenological expression adopted in
previous work does not represent a sensible approximation of this
intensity profile, and in fact leads to a gross underestimation of
the diffusion length.

In a subsequent work, we have shown that the relaxation of strain
at the outcrop of edge threading dislocations in GaN\{0001\} gives
rise to a piezoelectric field with a strength sufficient to dissociate
free excitons and to spatially separate electrons and holes at distances
over 100\,nm from the dislocation line \citep{kaganer18apl}. The
spatial separation inhibits radiative recombination of the electron-hole
pairs, and edge threading dislocations hence give rise to dark spots
in CL maps very similar to those observed experimentally even in the
absence of exciton diffusion. This result raises the question to what
extent the intensity contrast is actually still affected by exciton
diffusion, if at all.

Moreover, the strain field around a dislocation gives rise to a second
effect that so far has not been taken into consideration in the analysis
of the intensity contrast. The inhomogeneous strain field of a dislocation
\citep{hirthlothe82,IndenbomLothe} induces a change of the band gap
around the dislocation via the deformation potential mechanism \citep{BirPikus74,ghosh02,yan09,ishii10}.
The quasi-electric field of the band gap gradient around the dislocation
leads in turn to a drift of excitons in the strain field. Consequently,
the flux of excitons toward the dislocation, and thus the luminous
intensity due to their nonradiative annihilation at the dislocation,
is affected not only by diffusion but also by drift in the dislocation
strain field. The magnitudes of the drift and diffusion fluxes are
connected by the Einstein relation, but their driving forces are distinctly
different, and the directions of the fluxes may coincide or oppose
each other. In other words, exciton drift may enhance or counteract
the effects of exciton diffusion. In general, drift tends to concentrate,
whereas diffusion dissolves.

In the present work, we extend our previous theoretical framework
\citep{sabelfeld17CL} and explicitly consider, in addition to exciton
diffusion, the effects of the three-dimensional strain distribution
associated to an edge threading dislocation at the GaN\{0001\} surface,
i.\,e., exciton drift in the strain field and exciton dissociation
in the piezoelectric field induced by the changes in strain. We solve
this complex three-dimensional problem by an advanced Monte Carlo
scheme \citep{sabelfeld91book,sabelfeld16}. For comparison with the
theoretical predictions, we record hyperspectral CL maps around the
outcrop of a threading dislocation of a free-standing GaN(0001) film.
To facilitate the clear distinction of the effects of exciton drift
and diffusion, these maps are recorded for various temperatures ranging
from 10 to 200~K. Our measurements show that the CL \emph{intensity}
profiles do not notably depend on temperature, despite the fact that
the diffusion length is known to significantly decrease with increasing
temperature. In contrast, the CL \emph{energy} profile is observed
to depend strongly on temperature. These findings are reproduced by
our simulations. Our results show that the common understanding of
a diffusion-controlled intensity contrast around threading dislocations
in GaN\{0001\} is a misconception. The mechanism dominating the intensity
contrast is the piezoelectric field around the dislocation outcrop,
and exciton diffusion changes this contrast only marginally. However,
the energy contrast turns out to be highly sensitive to the diffusion
length, which we thus propose as a new experimental observable for
the actual determination of the carrier diffusion length in GaN.

\section{Methods}

\label{sec:methods}

The CL experiments were performed on the same free-standing GaN(0001)
layer grown by hydride vapor phase epitaxy as used in our previous
investigation \citep{sabelfeld17CL}. The density of threading dislocations
reaching the surface of this layer amounts to $6\times10^{5}$\,cm$^{-2}$.
CL spectroscopy was carried out using a Zeiss Ultra55 field-emission
scanning electron microscope equipped with a Gatan MonoCL4 system
and a He-cooling stage. The acceleration voltage was set to 3~kV,
and the probe current of the electron beam to 1.1~nA, with the aim
to minimize the generation volume of excitons as much as possible.
The electron beam diameter was 5~nm, significantly smaller than the step
size of 20~nm chosen for recording CL maps. 
Hyperspectral maps were acquired in the vicinity of isolated threading
dislocations for temperatures between 10 and 200\,K, using a parabolic
mirror to collect the emitted light, a spectrometer for dispersing
it with the spectral resolution set to 3\,meV, and a charge-coupled
device as the detector. Analysis of the CL data was performed using
the Python package HyperSpy \citep{hiperspy}.

The calculations of the CL intensity and energy around an edge threading
dislocation were performed using an advanced Monte Carlo scheme developed
previously for the solution of the pure diffusion problem \citep{sabelfeld91book},
but extended to include exciton drift in the strain field of the dislocation
\citep{sabelfeld16}, and exciton dissociation in the associated electric
field \citep{kaganer18apl}. The algorithm is described in detail
in the Supplementary Material. The material constants entering the
model (particularly the elastic moduli and the piezoelectric coefficients
$e_{31}$ and $e_{33}$ of GaN) were taken the same as in \citep{kaganer18apl}.
The value of the shear piezoelectric coefficient $e_{15}$ strongly
influences the CL intensity profiles \citep{kaganer18apl}, and was
here assumed to be $e_{15}=-e_{31}$. The surface recombination velocity
is taken to be $S=500$~nm/ns \citep{Aleksiejunas2003}. The diffusion
coefficient is assumed to be $D=2\times10^{5}$~nm$^{2}$/ns independent
of temperature since its temperature dependence is comparatively weak
\citep{scajev12}. Only one quantity entering the model, namely, the
exciton lifetime far from the dislocation, is treated as adjustable
parameter.

\section{The drift-diffusion-recombination problem}

\label{sec:equations}

We consider the generation of electron-hole pairs by an electron beam
incident on the surface of a GaN\{0001\} layer. In a CL or an EBIC
experiment, a tightly focused electron beam scans the sample and generates
electron-hole pairs. For the electron beam position $\mathbf{r}_e$,
the initial density of the electron-hole pairs generated by the electron
beam at a point $\mathbf{r}$ in the sample is described by a three-dimensional
generation function $Q(\mathbf{r}-\mathbf{r}_e)$. At low temperatures,
these electron-hole pairs rapidly bind to form excitons. At higher
temperatures, thermal dissociation results in the coexistence of excitons
and free carriers, with the ratio of their concentrations being controlled
by the Saha equation \citep{ebeling76}. Because of the exciton binding
energy of 26\,meV in GaN, and the comparatively high excitation density
in CL, the fraction of excitons remains large in the entire temperature
range of our experimental investigations. For simplicity, we therefore
refer exclusively to excitons in the following. It is important to
note, however, that the formalism we develop applies equally well
for minority carrier or ambipolar diffusion of electrons and holes\emph{,
}the only difference being the value of the diffusion coefficient.
The aim of this section is to formulate the problem of exciton diffusion,
drift, and recombination in the half-space containing a dislocation
line normal to the surface, with nonradiative annihilation of the
excitons at the free surface and in the vicinity of the dislocation
line.

Diffusion of excitons gives rise to the diffusional flux $-D\nabla n$,
where $D$ is the diffusion coefficient and $n(\mathbf{r})\equiv n(\mathbf{r};\mathbf{r}_e)$
is the three-dimensional distribution of the exciton density for a
given position $\mathbf{r}_e$ of the exciting electron beam. The spatial
variation of the band gap $E_{G}(\mathbf{r})$ due to the inhomogeneous
strain of the dislocation gives rise to the exciton drift velocity
$\mathbf{v}(\mathbf{r})=-\mu\nabla E_{G}$, where $\mu$ is the exciton
mobility, and to the drift flux $\mathbf{v}n$. The diffusion coefficient
$D$ and the mobility $\mu$ are related by the Einstein relation
$D=\mu kT$, where $k$ is the Boltzmann constant and $T$ is the
temperature.

\begin{figure}
\includegraphics[width=1\columnwidth]{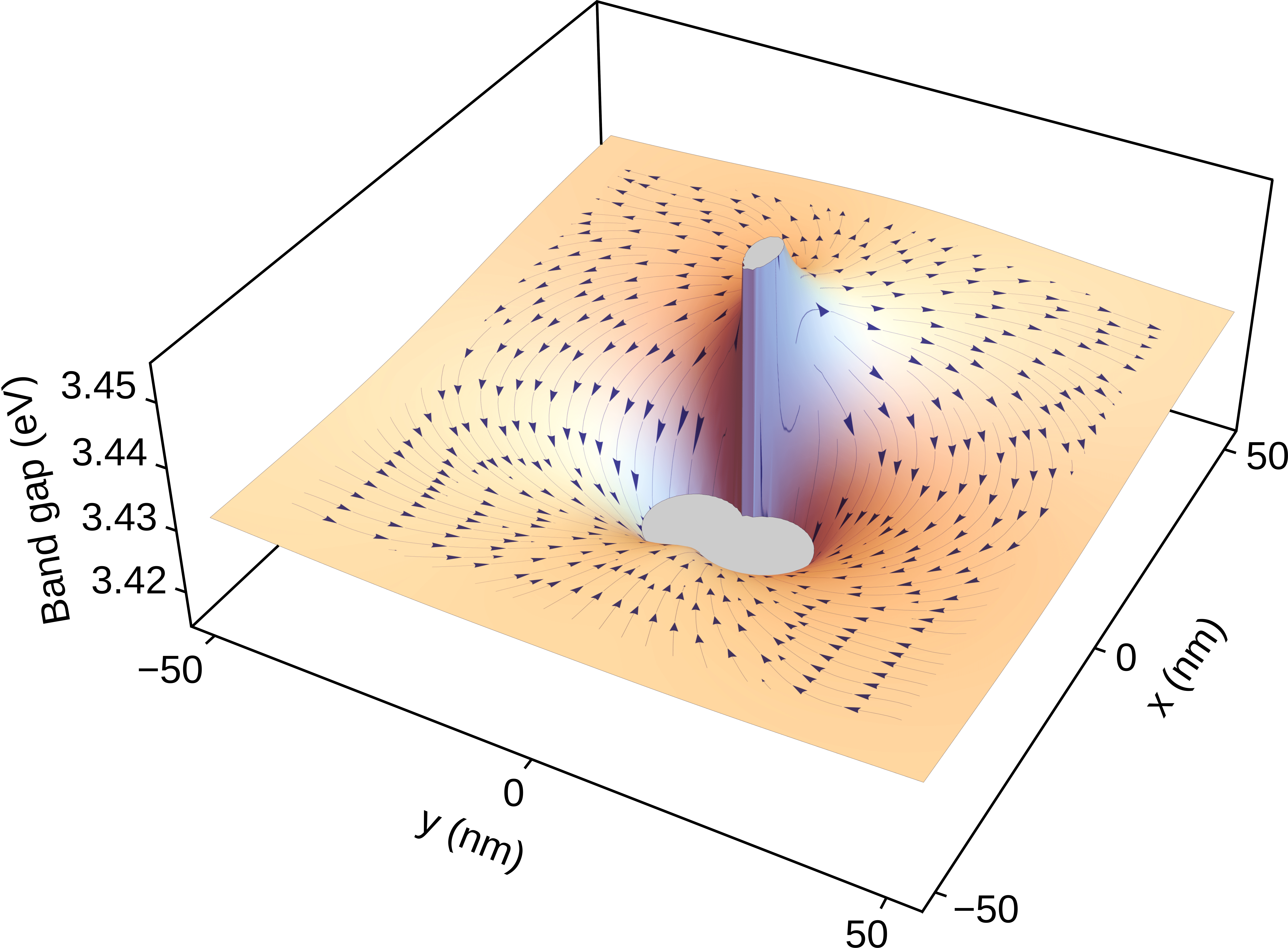}

\caption{Band-gap variation around the line of an \emph{a}-type edge dislocation
in GaN at room temperature and the drift velocity field caused by
this variation. The length of the arrow heads is proportional to the drift velocity.}

\label{fig:bandgap}
\end{figure}

The band gap $E_{G}(\mathbf{r})$ of a strained semiconductor generally
depends on all components of the strain tensor \citep{BirPikus74}.
However, the effect of shear strain is notably smaller than that of
normal strain. Since screw ($c$-type) threading dislocations induce
only shear strain, they have little effect on the band gap. Consequently,
we focus in this paper on the effect of the edge component of an edge
($a$-type) or mixed ($a+c$-type) threading dislocation. Figure \ref{fig:bandgap}
shows the variation of the band gap $E_{G}(\mathbf{r})$ of GaN in
the vicinity of an \emph{a}-type edge dislocation located at $x=y=0$
together with the respective drift velocity $\mathbf{v}(\mathbf{r})$,
calculated in the framework of a Bir-Pikus $\mathbf{k}\cdot\mathbf{p}$
approach \citep{ghosh02} taking into account both normal and shear
strain components. The shear strain gives rise to an asymmetry between
the regions of increased and decreased bandgap. The surface relaxation strain is not included in the calculation here, so that the band gap and the drift velocity shown in Fig.~\ref{fig:bandgap} are reached at the distances from the surface large compared to the respective lateral distance. 
Hence, the whole map corresponds to depths larger than 50~nm (while the central part is not affected by the surface relaxation strain also at smaller depths).
The band gap increases
in the region of compressive strain at one side and decreases in the
region of tensile strain at the other side of the dislocation line.
Due to the drift, excitons are hence repelled at one side of the dislocation
and attracted at the other side.

The total flux of excitons is the sum of the diffusion and the drift
fluxes,
\begin{equation}
\mathbf{J}=-D\nabla n+\mathbf{v}n,\label{eq:1}
\end{equation}
and the steady-state continuity equation reads

\begin{equation}
-\nabla\cdot\mathbf{J}-\frac{n}{\tau}+Q=0.\label{eq:2}
\end{equation}

In this equation, $\tau(\mathbf{r})$ is the exciton lifetime. It
contains contributions from the radiative lifetime $\tau_{r}$, the
nonradiative lifetime $\bar{\tau}_{nr}$ in the crystal far from dislocations,
and the position-dependent lifetime $\tau_{E}(\mathbf{r})$ describing
the rate of the exciton dissociation in the piezoelectric field near
the dislocation outcrop \citep{kaganer18apl}. Combining the rates
of the two nonradiative processes in one term, $1/\tau_{nr}(\mathbf{r})=1/\bar{\tau}_{nr}+1/\tau_{E}(\mathbf{r})$,
we can represent the total recombination rate as 
\begin{equation}
\frac{1}{\tau(\mathbf{r})}=\frac{1}{\tau_{r}}+\frac{1}{\tau_{nr}(\mathbf{r})}.\label{eq:2a}
\end{equation}
The drift-diffusion equation can thus be written as 
\begin{equation}
D\Delta n-\nabla\cdot(\mathbf{v}n)-\frac{n(\mathbf{r})}{\tau(\mathbf{r})}+Q(\mathbf{r}-\mathbf{r}_e)=0.\label{eq:3}
\end{equation}

Nonradiative exciton annihilation at the planar surface is described
by the boundary condition 
\begin{equation}
\left(-\mathbf{J}\cdot\boldsymbol{\nu}+Sn\right)\Bigr|_{\Gamma}=0,\label{eq:8}
\end{equation}
where $\Gamma$ is the surface $z=0$ of the half-infinite crystal,
$\boldsymbol{\nu}$ the outward surface normal vector, and $S$ the
surface recombination velocity. A similar boundary condition, with
an analogous recombination velocity, has been written at the dislocation
line represented by a thin cylinder \citep{sabelfeld17CL}.

Let us integrate Eq.~(\ref{eq:2}) over the sample volume. In the
first term of the equation, we proceed from a volume to a surface
integral, yielding the rate of exciton annihilation at the
surface $\Gamma$, 
\begin{equation}
I_{s}(\mathbf{r}_e)=\int_{\Gamma}\mathbf{J}\cdot\boldsymbol{\nu}\,d\sigma.\label{eq:9}
\end{equation}
This quantity can be directly measured in EBIC experiments. In the
second term of Eq.~(\ref{eq:2}), we separate the radiative and the
nonradiative contributions, which gives 
\begin{equation}
I_{s}(\mathbf{r}_e)+\frac{1}{\tau_{r}}\int n(\mathbf{r};\mathbf{r}_e)\,d\mathbf{r}+\int\frac{n(\mathbf{r};\mathbf{r}_e)}{\tau_{nr}(\mathbf{r})}\,d\mathbf{r}=\int Q(\mathbf{r})\,d\mathbf{r}.\label{eq:9a}
\end{equation}
The right hand side of this equation is the total number of excitons
generated per unit time. The terms in the left hand side are the flux
to the surface $I_{s}(\mathbf{r}_e)$ and the number of excitons recombining
per unit time radiatively and nonradiatively, respectively. The total photon flux
\begin{equation}
I_{\mathrm{CL}}(\mathbf{r}_e)=\tau_{r}^{-1}\int n(\mathbf{r};\mathbf{r}_e)\,d\mathbf{r}\label{eq:9b}
\end{equation}
is the quantity measured in a CL experiment, and our primary aim is
its calculation.

An exciton radiatively recombining at some point $\mathbf{r}$ in
the sample produces a photon whose energy is equal to $E_{G}(\mathbf{r})-E_{X}$,
where $E_{G}(\mathbf{r})$ is the local bandgap and $E_{X}$ is the
exciton binding energy. Depending on the position $\mathbf{r}$ of
the electron beam with respect to the dislocation, the photons are
predominantly generated in the regions of compression or tension in
the dislocation strain field, giving rise to blue and red shifts of
the exciton line, respectively \citep{gmeinwieser05,liu16}. The exciton
line position is given by averaging the photon energy $E_{G}(\mathbf{r})-E_{X}$
over the distribution of excitons recombining radiatively. This conditional
expectation is 
\begin{equation}
E_{CL}(\mathbf{r}_e)=\frac{\tau_{r}^{-1}\int\left[E_{G}(\mathbf{r})-E_{X}\right]n(\mathbf{r};\mathbf{r}_e)\,d\mathbf{r}}{\tau_{r}^{-1}\int n(\mathbf{r};\mathbf{r}_e)\,d\mathbf{r}},\label{eq:9c}
\end{equation}
and it is calculated in the present work together with the CL intensity
distribution, and compared with the experimental results.

The drift-diffusion problem could in principle be solved by a direct
Monte Carlo modeling of random walks on a grid performed by particles
generated from the source, with a finite lifetime and a partial annihilation
at the boundaries. Such modeling was performed, for pure diffusion
with constant lifetime and neglecting drift, in Refs.~\citep{tabet98ssp,tabet98sst,ledra05}.
However, a direct modeling of a three-dimensional random walk is extremely
time-consuming, since the particle makes many random steps before
it annihilates in the bulk or meets a boundary.

An efficient Monte Carlo algorithm developed for the diffusion problem
\citep{sabelfeld91book} and generalized recently to the drift-diffusion
problem \citep{sabelfeld16} replaces the random walk inside a sphere
by a single, appropriately modeled, step from the center of the sphere
to its surface. The sphere radius can be taken equal to the distance
to the nearest surface. The survival probability for the particle
to reach the sphere, depending on its finite lifetime, is explicitly
calculated. If the particle survives, it jumps to a random point on
the sphere, and the next sphere is generated with its radius given
by the distance from the new position to the nearest surface. It has
been shown that the sequence of spheres converges to the boundary.
The mean number of steps required to reach a distance $\varepsilon$
from the boundary is proportional to $|\log\varepsilon|$, so that
even for a very small value of $\varepsilon$ the mean number of steps
in this process is quite small. The behavior of the particle after
it hits the $\varepsilon$-vicinity of the surface is described separately,
in accordance with the given boundary conditions. A detailed account
of the Monte Carlo algorithm for the solution of the drift-diffusion
problem can be found in the Supplementary Material.

\section{Results}

\label{sec:results}

\begin{figure}
\includegraphics[width=0.8\columnwidth]{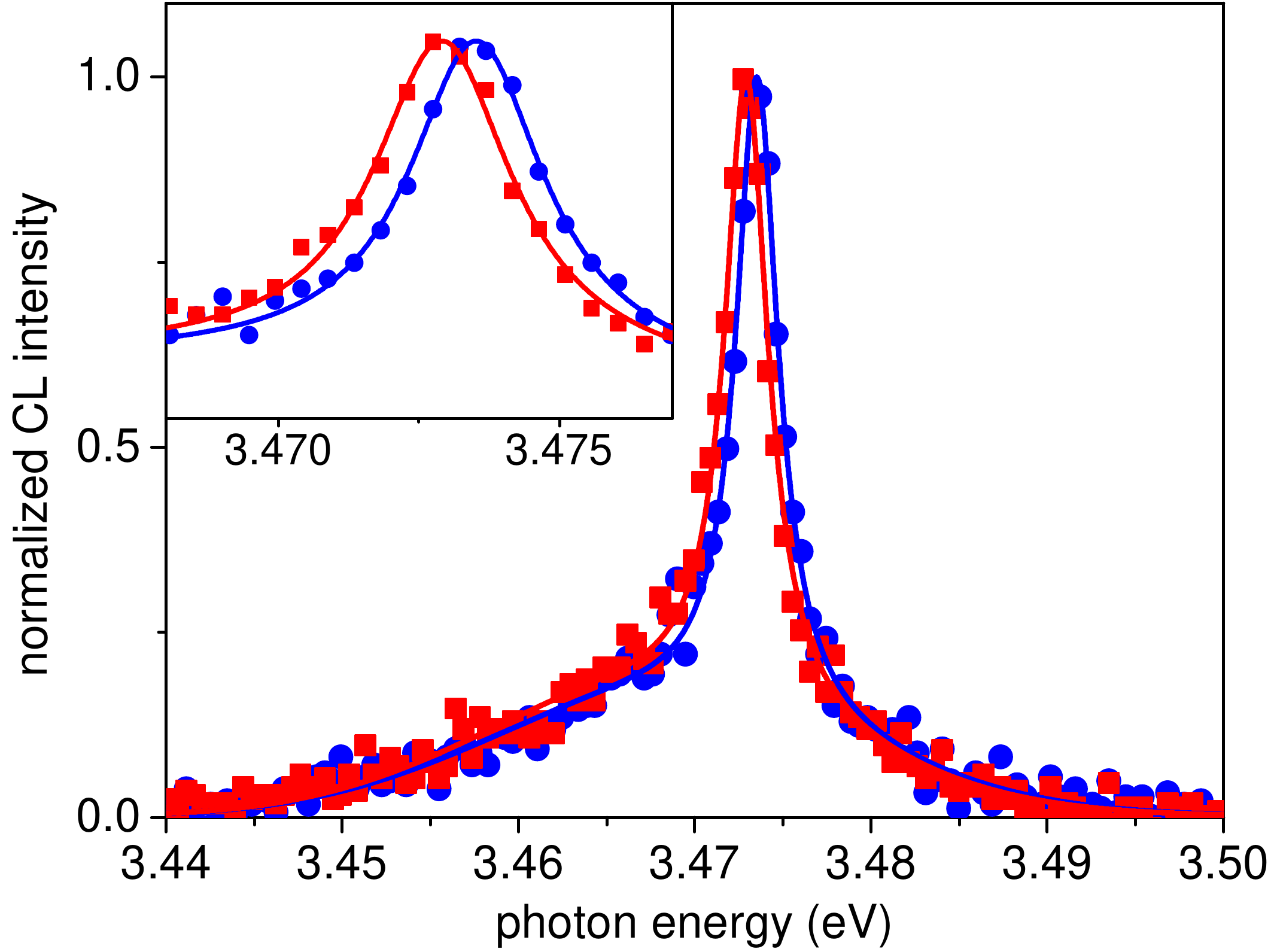}

\caption{Experimental CL spectra measured at 10~K at the positions corresponding
to the maximum blue (circles) and red shift (squares) of the exciton
transitions in the vicinity of a selected threading dislocation in
our GaN(0001) layer. The data are fit by a line-shape model (lines)
taking into account the different excitonic contributions. The inset
shows an expanded view of the spectra.}

\label{fig:spectra}
\end{figure}

For recording hyperspectral maps around the outcrop of threading dislocations,
the electron beam was raster-scanned across the sample surface of
$2\times2$\,\textmu m$^{2}$ containing the dislocation of interest
with a sampling step size, i.\,e., a spatial pixel, of 20\,nm. At
every step, a full CL spectrum was recorded in the energy range from
3.2 to 3.8\,eV, covering all free and bound exciton transitions for
our free-standing GaN layer \cite{monemar08}. We have investigated a significant number ($>10$) of dislocations for this sample, and the results shown below are representative for all dislocations and vary within $\pm 20$\%.

Figure \ref{fig:spectra} shows two exemplary
CL spectra taken from such a map recorded around a selected threading
dislocation at 10~K. The spectra correspond to the points in the
maps exhibiting the maximum blue and red shift close to the outcrop
of the threading dislocation. Each spectrum was fit by the sum of
a Lorentzian and a Gaussian accounting for the dominant high-energy
line stemming from donor-bound exciton recombination \cite{freitas02}, and the low-energy
shoulder originating from acceptor-bound excitons as well as from
the two electron satellites of the donor-bound exciton \cite{wysmolek02}, respectively,
as shown in Fig.\,\ref{fig:spectra} (for maps acquired at 50, 120,
and 200\,K, no bound exciton transition is observed and the fits
to the free \emph{A} exciton line \cite{monemar96,kovalev96} were done with a single Lorentzian).

The sharp lines in Fig.~\ref{fig:spectra} are resolution limited, with the full width at half maximum (FWHM) of 3~meV. The accuracy of the spectral positions of the lines is governed by statistical fluctuations of the CL intensity and amounts to 0.04~meV, as obtained from the fits shown in Fig.~\ref{fig:spectra}. The lines broaden with increasing temperature and reach a FWHM of 25~meV at 200~K. The accuracy than reduces to 0.4~meV, because of larger statistical fluctuations of the intensity. Since the lineshifts in the vicinity of the dislocation also increase with temperature, as it is demonstrated below, the accuracy is still sufficient to systematically study both CL intensity and line positions. However, the measurements at temperatures above 200~K are too noisy for a reliable analysis. 

\begin{figure*}
\includegraphics[width=0.855\textwidth]{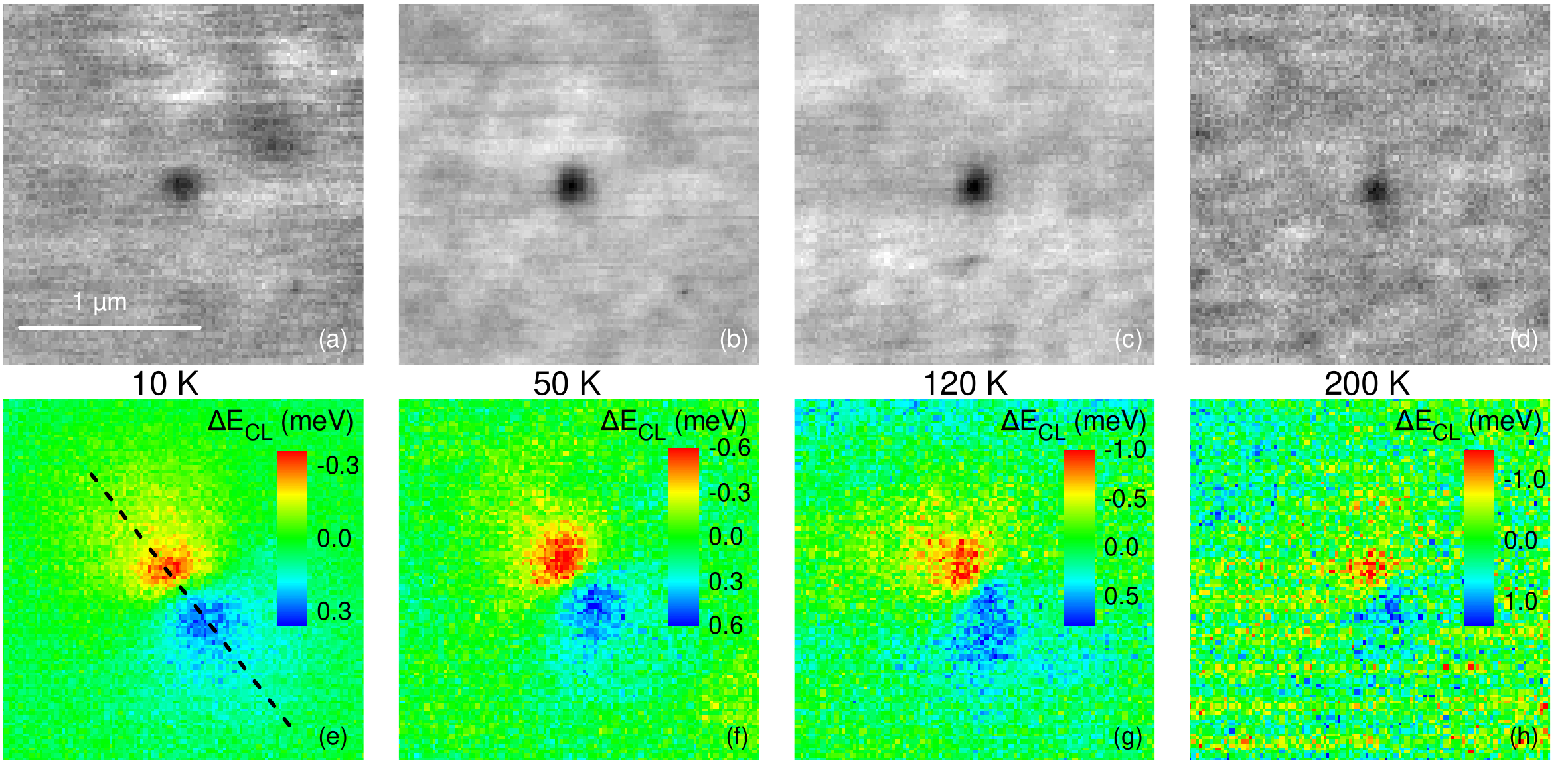}

\caption{Two-dimensional spatial maps of the (a--d) spectrally integrated
CL intensity and (e--h) CL spectral line positions for temperatures
between 10 and 200\,K around the selected threading dislocation.
The dashed line in (e) indicates the direction of the line scans shown
in Figs.~\ref{fig:curves}. The scale bar in (a) applies to all maps.}

\label{fig:maps}
\end{figure*}

For each temperature, the fits yield maps of the spectrally integrated
intensity of the dominant exciton transition and its spectral position,
as shown in Figs.~\ref{fig:maps}(a--d) and Figs.~\ref{fig:maps}(e--h),
respectively. The maps clearly reveal the reduced CL intensity and
the shift of the transition energy at the outcrop of the threading
dislocation. Note that the maps were taken from one and the same dislocation
for all temperatures.

The red and the blue lobes in the maps in Figs.~\ref{fig:maps}(e--h)
correspond to regions of tensile and compressive strain around the
dislocation, respectively, and hence demonstrate that this dislocation
is of either $a$ or $a+c$ type. Since the screw component induces
only shear strain with a minor effect on the band gap, we cannot distinguish
pure edge and mixed dislocations. The line shift increases with temperature
and reaches 2.5~meV at 200~K. At the same time, the statistical
error of the fits increases due to the thermal quenching of the emission
intensity and the resulting increasing noise in the spectra, thus
leading to a higher noise level also in the maps.

\begin{figure}
\includegraphics[width=1\columnwidth]{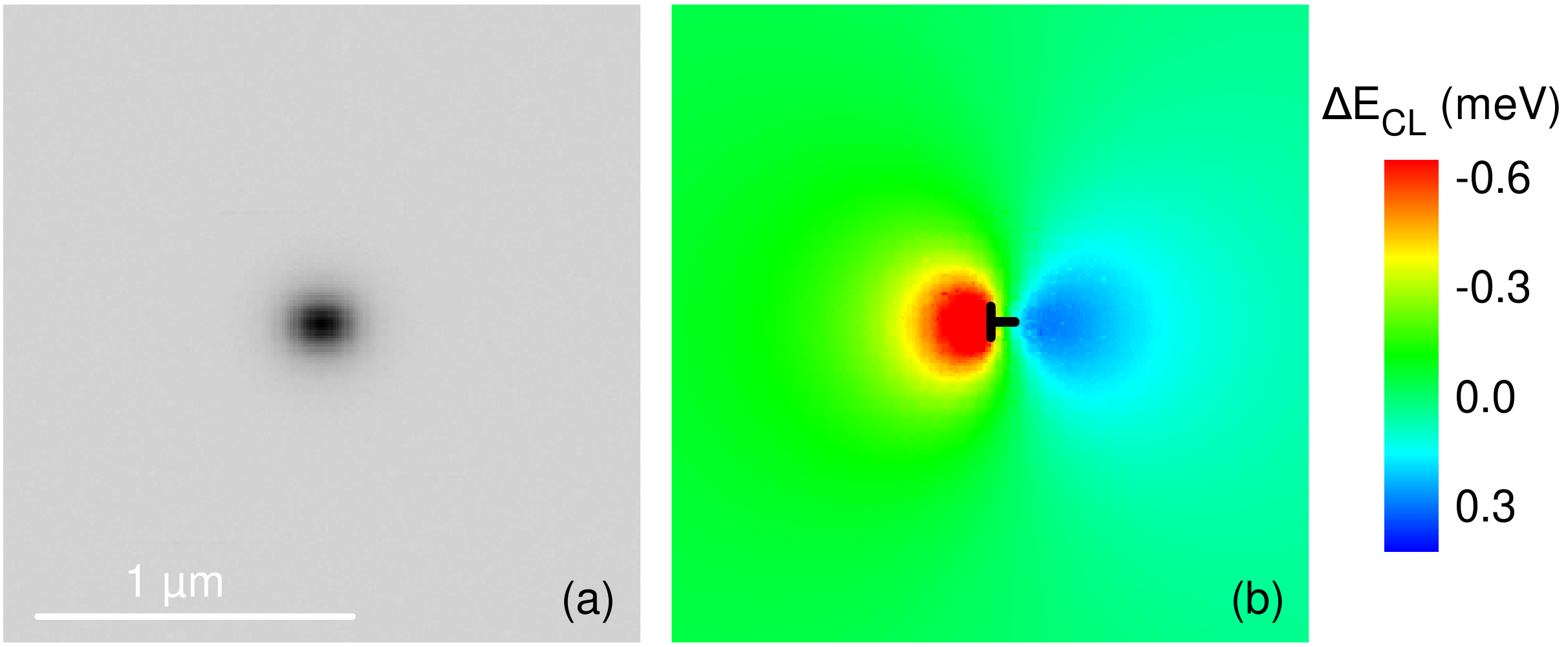}

\caption{Monte Carlo calculated maps of (a) spectrally integrated CL intensity
and (b) CL spectral line positions for a diffusion length $L=100$~nm
and a temperature of 100~K. The extra half-plane of the dislocation
indicates the dislocation position in (b).}

\label{fig:MCmaps}
\end{figure}

Figures\,\ref{fig:MCmaps}(a,b) show simulated maps of the spectrally
integrated CL intensity and CL spectral line positions around an edge
threading dislocation, respectively. Evidently, the maps are close
to the experimentally recorded maps shown in Fig.\,\ref{fig:maps}.
In the CL intensity map, the dislocation outcrop is associated with
a dark spot with FWHM of about 200~nm.
The intensity distribution appears to be almost isotropic, i.\,e.,
the anisotropy of the piezoelectric field around the dislocation \citep{kaganer18apl}
is effectively smoothed out by exciton diffusion. The change in the
CL line position is seen at distances up to 1~\textmu m from the
dislocation. The cut of an extra plane at the dislocation line indicated
in Fig.~\ref{fig:MCmaps}(b) shows the dislocation position at $x=0$. One can
see that a vertical line separating the blue and red lobes does not pass through the dislocation position but is shifted with respect to it. Lattice compression in the right part
gives rise to a blue shift, while lattice expansion in the left part
induces a red shift.

\begin{figure*}
\includegraphics[width=0.7\textwidth]{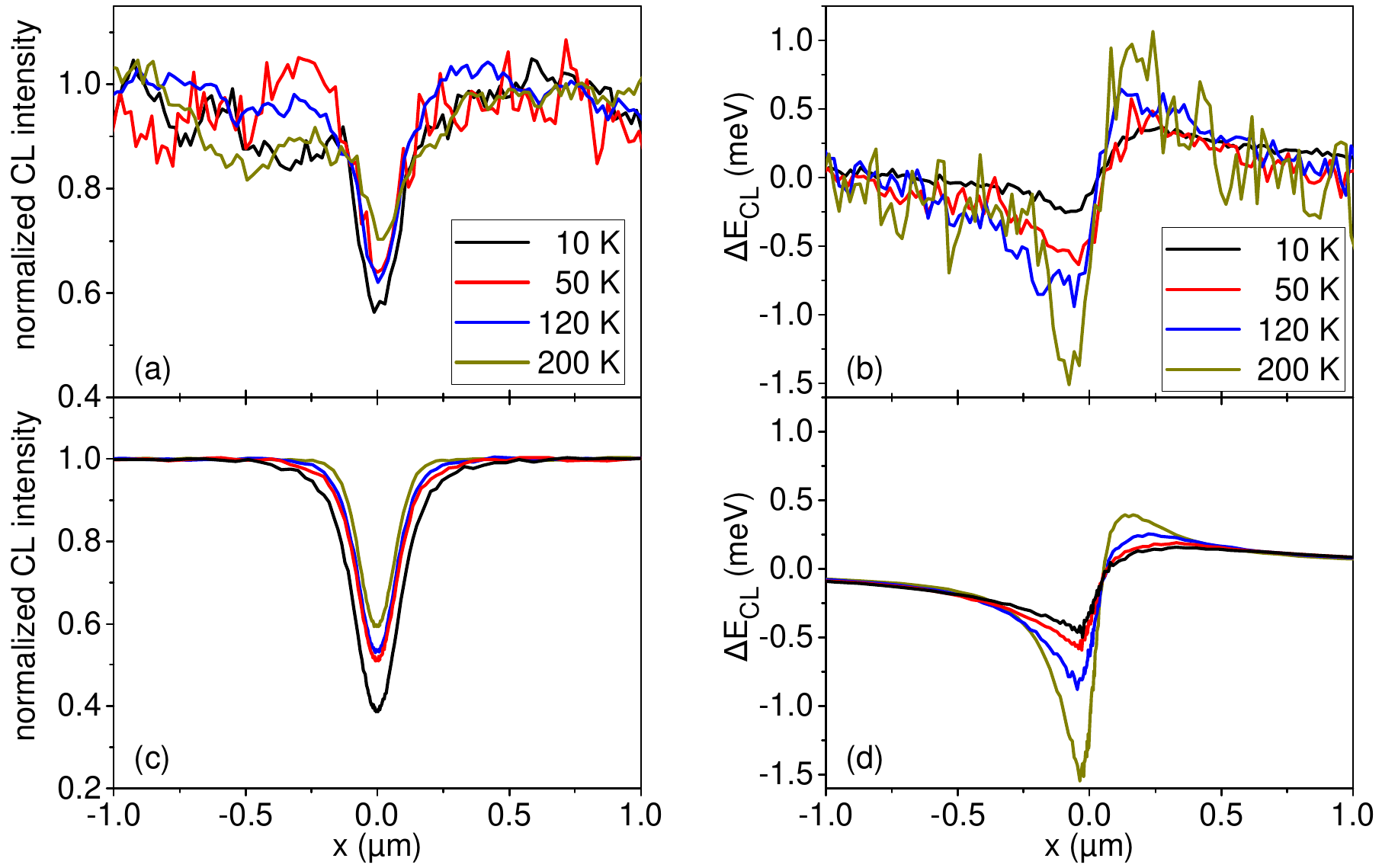}

\caption{Line profiles along the strain dipole of the dislocation {[}dashed
line in Fig.~\ref{fig:maps}(e){]}. (a) Experimental and (c) simulated
spectrally integrated CL intensity. (b) Experimental and (d) simulated
spectral position of the dominant exciton line.}

\label{fig:curves}
\end{figure*}

For a quantitative analysis, Figs.\,\ref{fig:curves}(a,b) show the
integrated intensity and the spectral position of the exciton transition
obtained by line scans across the strain dipole of the dislocation
{[}dashed line in Fig.~\ref{fig:maps}(e){]} in Figs.~\ref{fig:maps}(a--h).
Figure \ref{fig:curves}(a) clearly shows that the width of the CL
intensity curves does not notably depend on temperature, although
the diffusion length is expected to vary strongly in this temperature
range \citep{scajev12}. In contrast, the amplitude of the energy
contrast across the dislocation shown in Fig.~\ref{fig:curves}(b)
clearly increases with increasing temperature.

Figures \ref{fig:curves}(c,d) present the results of the corresponding
Monte Carlo simulations, with the only free parameter entering the
calculations, namely, the effective exciton lifetime in the crystal
far from the dislocation $\tau_{0}=\tau(|\mathbf{r}|\rightarrow\infty)$
{[}cf.~Eq.~(\ref{eq:2a}){]}, chosen such as to reproduce the experimental
trends, especially the energy profiles in Fig.\,\ref{fig:curves}(b).
Evidently, a variation of the lifetime alone is sufficient to reproduce
the trend observed for both the intensity {[}Fig.\,\ref{fig:curves}(c){]}
and energy profiles {[}Fig.\,\ref{fig:curves}(d){]}. We note that
we assumed an actual temperature of 20\,K for the calculated profiles
at a nominal temperature of 10\,K, since the intensity contrast is
otherwise much stronger than observed experimentally due to exciton
drift. In fact, the carrier temperature deduced from the high-energy
slope of the 10\,K-CL spectrum is 30\,K. We also note that we could
have easily improved the agreement between experiment and theory by
adjusting, for example, the surface recombination velocity, the piezoelectric
coefficient $e_{15}$, or the partial screening of the piezoelectric
field by a moderate ($10^{16}$\,cm$^{-3}$) background doping \citep{kaganer18apl}.

\begin{figure}
\includegraphics[width=0.6\columnwidth]{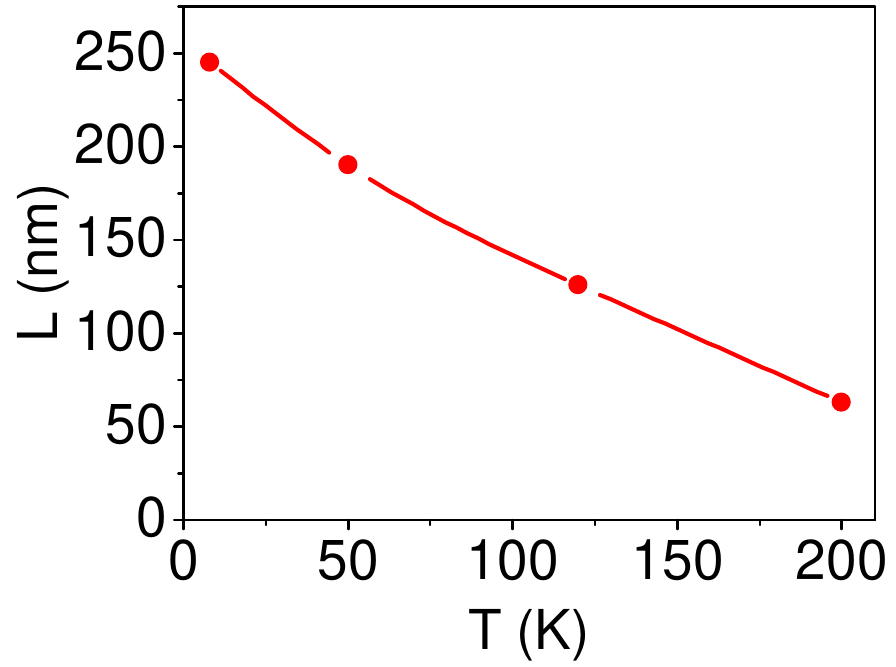}

\caption{Temperature dependence of the exciton diffusion length obtained from
the simulations in Fig.~\ref{fig:curves}(c,d).}

\label{fig:Ldiff}
\end{figure}

Since we assume a constant diffusion coefficient $D=2\times10^{5}$~nm$^{2}$/ns,
the temperature dependence of the diffusion length $L=\sqrt{D\tau_{\mathrm{0}}}$
presented in Fig.~\ref{fig:Ldiff} is determined by the temperature
dependence of the effective exciton lifetime $\tau_{\mathrm{0}}$
far from the dislocation. $L$ is seen to vary by a factor of five
from 10 to 200\,K, while the width of the intensity profiles changes
in the same temperature range by only 10\%. In fact, when analyzing
these profiles by our previous model, in which the dislocation line
acts as the only nonradiative center for excitons, an essentially
temperature-independent, but much larger value of the diffusion length
is obtained \citep{sabelfeld17CL}. For example, a diffusion length
of 400\,nm was obtained in Ref.\,\citep{sabelfeld17CL} from the
room-temperature profile taken on the same sample as in the present
work.

\begin{figure*}
\includegraphics[width=1\textwidth]{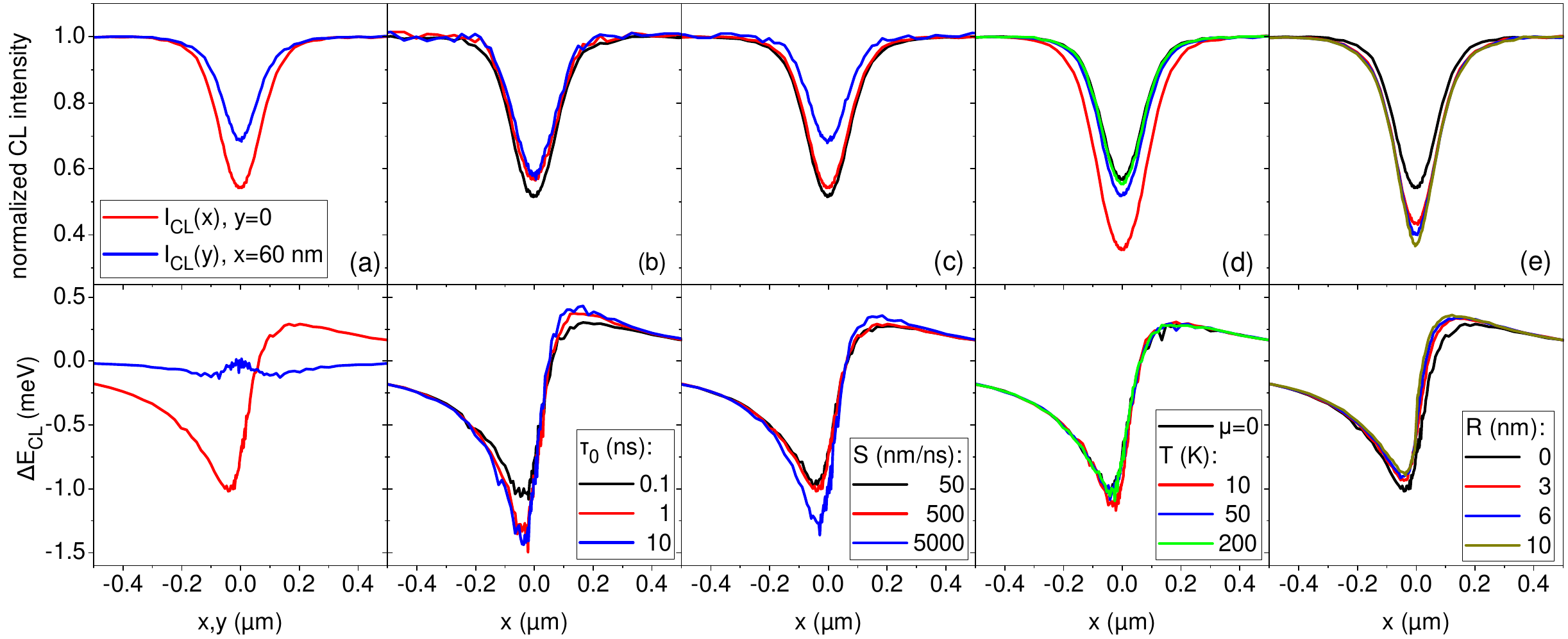}

\caption{Monte Carlo simulations of the normalized CL intensity (top) and the
average CL line shift (bottom) across an edge threading dislocation
in GaN(0001). (a) Intensity and energy profiles for $T=100$~K with
a surface recombination velocity $S=500$~nm/ns and a diffusion length
$L=100$~nm. The scans are taken along the $x$ axis at $y=0$ and
along the $y$ axis with an offset $x=60$~nm, along the line with
the minimum energy contrast variation. (b--d) Profiles along the
$x$ axis obtained when varying (b) the exciton lifetime $\tau_{0}$ far
from the dislocation while keeping a constant diffusion length $L=100$~nm, (c)
the surface recombination velocity $S$, and (d) the temperature $T$
(and thus the impact of exciton drift), respectively. (e) Profiles
obtained when adding a nonradiative dislocation line represented by
a cylinder of radius $R$ and infinite surface recombination velocity.}

\label{fig:simulations}
\end{figure*}

We finally examine the impact of the various parameters entering the
problem. Figure~\ref{fig:simulations}(a) shows the simulated intensity
and energy profiles extracted from the maps presented in Fig.~\ref{fig:MCmaps}
both along $(x)$ and across $(y)$ to the strain dipole. We have already seen in Fig.~\ref{fig:MCmaps}(b) that the line separating the blue and red lobes is shifted with respect to the dislocation position, as indicated in Fig.~\ref{fig:MCmaps}(b) by the cut of the extra plane.
This effect is also evident in both experimental and simulated profiles in Fig.~\ref{fig:curves}:
at the position of the dislocation at $x=0$, the energy is strongly
red-shifted, while a zero energy shift is observed at $x\approx60$~nm.
Accordingly, the scan along the $y$ axis in Fig.~\ref{fig:simulations}(a)
is taken at $x=60$~nm.

Figures\,\ref{fig:simulations}(b--d) show simulated intensity and
energy profiles obtained when changing one parameter at a time compared
to the profiles in Fig.\,\ref{fig:simulations}(a). In Fig.\,\ref{fig:simulations}(b),
we vary the exciton lifetime $\tau_{0}$ far from the dislocation
while keeping the diffusion length $L=\sqrt{D\tau_{0}}=100$~nm constant
by simultaneously varying the diffusion coefficient $D=L^{2}/\tau_{0}$.
Lifetimes of 1\,ns and longer reduce the CL intensity and enhance
the CL energy contrast, but the effect is weak considering the order
of magnitude variation of the lifetime. Figure \ref{fig:simulations}(c)
shows the effect of surface recombination. A decrease of the surface
recombination velocity does not notably change either intensity or
energy profile, while its increase reduces the intensity and enhances
the energy contrast. Figure \ref{fig:simulations}(d) demonstrates
the effect of drift, assuming that all other parameters do not change
with temperature. The relative effects of drift and diffusion depend
on temperature, since the mobility and the diffusivity are connected
by the Einstein relation $D=\mu kT$. Our simulations show that the
energy profile is hardly affected by drift regardless of the temperature.
The intensity contrast is strongly enhanced at low temperatures, while
the profile for $T>200$\,K already approaches the high-temperature
limit ($\mu=0$).

Finally, for the simulations shown in Figs.~\ref{fig:curves}(c,d)
and \ref{fig:simulations}(a--d), we have assumed that the only effect
of the dislocation is exciton dissociation in the piezoelectric field
around the dislocation outcrop \citep{kaganer18apl}. This assumption
is a radical change of the existing paradigm, within which the dark
spots in CL maps are related to the nonradiative character of the
dislocation line itself, caused by either dangling bonds or point
defects accumulating around the line. In theoretical studies, dislocations
are thus usually modeled as thin cylinders acting as sinks for carriers
\citep{lax78,donolato78,donolato79,donolato85,donolato98,jakubowicz85,jakubowicz86,pasemann91,sabelfeld17CL}.
In Fig.~\ref{fig:simulations}(e), we add such a cylinder with radius
$R$ in the simulations, assuming an infinite recombination velocity
at the surface giving rise to the total nonradiative annihilation
of excitons. This worst-case assumption results in an increase of
the intensity contrast, but has only a minor effect on the energy
contrast. The reason for this only marginal impact of the dislocation
line is the fact that most carriers are generated close to the surface
(and thus in the reach of the piezofield) for an acceleration voltage
of 3\,kV.

\section{Summary and Conclusions}

An edge or mixed threading dislocation at the GaN\{0001\} surface
is surrounded by a strong piezoelectric field that dissociates and
spatially separates electron hole pairs, thus inhibiting their radiative
recombination. The characteristic size of the region of reduced luminous
intensity is determined by the spatial extent of the field, and is
insensitive to the exciton diffusion length. The key for understanding
this insensitivity is the fact that excitons generated within the
reach of the piezoelectric field have no chance to escape its influence,
since their lifetime (and hence diffusion length) is reduced to effectively
zero regardless of their diffusion length far from the dislocation.
In fact, the finite contrast observed for excitation within the \emph{lateral}
reach of the piezoelectric field stems from excitons generated beneath
its reach \emph{in depth}. This fraction of excitons excited deeper
in the crystal also dominates the spectral position of the exciton
transition close to the dislocation line. For short diffusion lengths,
excitons decay radiatively close to their point of excitation and
thus probe the high strain close to the dislocation line. In contrast,
long diffusion lengths lead to a spatial redistribution of excitons,
the majority of which will then decay far from the dislocation line
with transition energies close to the bulk value, hence smoothing
out the characteristic variation of the energy position in the vicinity
of a dislocation. Energy profiles across edge dislocation outcrops
at the GaN\{0001\} surface are thus a sensitive means to access the
exciton or carrier diffusion length in GaN, while intensity profiles
are not.
\begin{acknowledgments}
The authors thank Henning Riechert and Uwe Jahn for a critical reading
of the manuscript. The free-standing GaN layer is courtesy of Ke Xu
and Hui Yang from the Suzhou Institute of Nano-Tech and Nano-Bionics.
The Monte Carlo simulations presented in the paper were performed
at the Siberian Supercomputer Center of the Siberian Branch of the
Russian Academy of Sciences. K.K.S. and A.E.K. acknowledge the support
of the Russian Science Foundation under Grant No. 19-11-00019. 
\end{acknowledgments}

\newpage{}

\section*{Supplementary material}

\setcounter{subsection}{0}

\title{Supplementary material to the paper: \\
"Determination of the carrier diffusion length in GaN from cathodoluminescence maps around threading dislocations: fallacies and opportunities"}
\author{Vladimir M. Kaganer}
\affiliation{Paul-Drude-Institut f\"ur Festk\"orperelektronik, Leibniz-Institut
im Forschungsverbund Berlin e.\,V., Hausvogteiplatz 5--7, 10117
Berlin, Germany}
\author{Jonas L\"ahnemann}
\affiliation{Paul-Drude-Institut f\"ur Festk\"orperelektronik, Leibniz-Institut
im Forschungsverbund Berlin e.\,V., Hausvogteiplatz 5--7, 10117
Berlin, Germany}
\author{Carsten Pf\"uller}
\affiliation{Paul-Drude-Institut f\"ur Festk\"orperelektronik, Leibniz-Institut
im Forschungsverbund Berlin e.\,V., Hausvogteiplatz 5--7, 10117
Berlin, Germany}
\author{Karl K. Sabelfeld}
\affiliation{Institute of Computational Mathematics and Mathematical Geophysics,
Russian Academy of Sciences, Lavrentiev Prosp.\ 6, 630090 Novosibirsk,
Russia}
\author{Anastasya E. Kireeva}
\affiliation{Institute of Computational Mathematics and Mathematical Geophysics,
Russian Academy of Sciences, Lavrentiev Prosp.\ 6, 630090 Novosibirsk,
Russia}
\author{Oliver Brandt}
\affiliation{Paul-Drude-Institut f\"ur Festk\"orperelektronik, Leibniz-Institut
im Forschungsverbund Berlin e.\,V., Hausvogteiplatz 5--7, 10117
Berlin, Germany}

\maketitle

In this Supplement, we present a derivation and a detailed description
of the Monte Carlo algorithm for solving the drift-diffusion problem
formulated in Sec.~III of the paper. References
to equations in the main text of the paper are marked by asterisks.

\subsection{Reciprocity theorem}

A direct implementation of the drift-diffusion problem  requires us to first solve the drift-diffusion equation (4){*}
containing the source $Q(\mathbf{r})$ with the boundary conditions
(5){*}, and then to integrate the solution over the sample
volume by Eq.~(8){*} to obtain the CL intensity. The calculation
of the total flux to the surface by Eq.~(6){*}, to get
the EBIC intensity, requires an integration over the surface. However,
we show that there is another, more direct approach, in which the
CL intensity $I_{\mathrm{CL}}$ and the flux to the surface $I_{s}$
are obtained as solutions of certain adjoint boundary value problems.
The basis of this approach is the reciprocity theorem, which was initially
formulated for the flux $I_{s}$ in the case of pure diffusion and
an infinite surface recombination velocity \citep{donolato85}. We
have recently extended this theorem to arbitrary recombination velocities
at all surfaces \citep{sabelfeld17CL}. Here we generalize the reciprocity
relation to the case of the drift-diffusion equation, both for $I_{s}$
and $I_{\mathrm{CL}}$.

Let us consider Eq.~(4){*} with a steady state unit 
point source at $\mathbf{r}_e$. Namely, let us introduce an arbitrary unit time $\tau_0$ and consider the source 
$Q(\mathbf{r})=\tau_0^{-1} \delta(\mathbf{r}-\mathbf{r}_e)$. The solution
of this equation $n(\mathbf{r})$ with the boundary condition (5){*}
gives, after integration over the sample volume (8){*},
the CL intensity $I_{\mathrm{CL}}(\mathbf{r}_e)$. Consider
now the solution $w(\mathbf{r})$ of the boundary value problem 
\begin{equation}
D\Delta w+\mathbf{v}\cdot\nabla w-\frac{w}{\tau}+\frac{1}{\tau_0}=0 \label{eq:16:prime}
\end{equation}
with the boundary condition 
\begin{equation}
\left(D\nabla w\cdot\boldsymbol{\nu}+Sw\right)\Bigr|_{\Gamma}=0~.\label{eq:12:prime}
\end{equation}
According to the adjoint  equation (\ref{eq:16:prime}),  $w(\mathbf{r})$ is the number of excitons in the  point $\mathbf{r}$ when a stationary source generates excitons homogeneously in the bulk with a unit rate.

We can prove that 
\begin{equation}
w(\mathbf{r}_e)=\int n(\mathbf{r})\,d\mathbf{r}, \label{eq:12b}
\end{equation}
and hence $I_{\mathrm{CL}}(\mathbf{r}_e)=\tau_r^{-1} w(\mathbf{r}_e)$. For
that purpose, multiply the direct equation (4){*} {[}with
the point source in it, $Q(\mathbf{r})=\tau_0^{-1}\delta(\mathbf{r}-\mathbf{r}_e)${]}
with $w$, the adjoint equation (\ref{eq:16:prime}) with $n$, and
subtract. The resulting equation can be written as 
\begin{equation}
\nabla\cdot\left[D\left(w\nabla n-n\nabla w\right)-\mathbf{v}nw\right]=-\frac{w}{\tau_0}\delta(\mathbf{r}-\mathbf{r}_e)+\frac{n}{\tau_{0}} \,.\label{eq:12a}
\end{equation}
Now we integrate this equation over the volume and apply the Green
formula. It gives 
\begin{eqnarray}
\int\limits _{\Gamma}\left[D\left(w\nabla n-n\nabla w\right)-\mathbf{v}nw\right]\cdot\boldsymbol{\nu}d\sigma \\ \nonumber
=-\frac{w(\mathbf{r}_e)}{\tau_0}+\frac{1}{\tau_{0}}\int n(\mathbf{r})\,d\mathbf{r}.\label{eq:141}
\end{eqnarray}
The left-hand side of this equation is equal to zero due to the boundary
conditions (5){*} and (\ref{eq:12:prime}), and we arrive at Eq.~{eq:12b}, which completes the proof.

A similar reciprocity relation can be formulated for the flux to the
surface $I_{s}(\mathbf{r}_e)$. Namely, consider the solution
$\tilde{w}(\mathbf{r})$ of the homogeneous equation
\begin{equation}
D\Delta\tilde{w}+\mathbf{v}\cdot\nabla\tilde{w}-\frac{\tilde{w}}{\tau}=0\label{eq:16}
\end{equation}
with the modified boundary condition 
\begin{equation}
\left(D\nabla\tilde{w}\cdot\boldsymbol{\nu}+S\tilde{w}\right)\Bigr|_{\Gamma}=\frac{S}{\tau_0}.\label{eq:12}
\end{equation}
Now the source in the adjoint equation is on the boundary and produces excitons homogeneously over the boundary.

Then, we can show that $I_{s}(\mathbf{r}_e)=\tilde{w}(\mathbf{r}_e)$.
The proof is similar to the one above. Let us multiply the direct
equation (4){*} {[}with the point source in it, $Q(\mathbf{r})=\tau_0^{-1}\delta(\mathbf{r}-\mathbf{r}_e)${]}
with $\tilde{w}$, the adjoint equation (\ref{eq:16}) with $n$,
and subtract. The resulting equation can be written as 
\begin{equation}
\nabla\cdot\left[D\left(\tilde{w}\nabla n-n\nabla\tilde{w}\right)-\mathbf{v}n\tilde{w}\right]=-\frac{\tilde{w}}{\tau_0}\delta(\mathbf{r}-\mathbf{r}_e).\label{eq:13}
\end{equation}
Now we integrate this equation over the volume and apply the Green
formula. This gives 
\begin{equation}
\int_{\Gamma}\left[D\left(\tilde{w}\nabla n-n\nabla\tilde{w}\right)-\mathbf{v}n\tilde{w}\right]\cdot\boldsymbol{\nu}d\sigma=-\frac{\tilde{w}(\mathbf{r}_e)}{\tau_0}.\label{eq:14}
\end{equation}
At the boundary $\Gamma$, the integrand is equal to $-Sn(\mathbf{r})/\tau_0$,
and using the boundary condition (5){*} once again, we arrive
at 
\begin{equation}
\tilde{w}(\mathbf{r}_e)=\int_{\Gamma}\mathbf{J}\cdot\boldsymbol{\nu}d\sigma,\label{eq:15}
\end{equation}
i.e., the solution of the homogeneous equation (\ref{eq:16}) with
the boundary condition (\ref{eq:12}) is equal to the total flux per unit time to
the surface $I_{s}(\mathbf{r}_e)$ defined by Eq.~(6){*}.

It is worth to note that similar reciprocity relations can be formulated
for problems that involve several surfaces with different surface
recombination velocities, particularly when the dislocation is considered
as a thin cylinder characterized by its surface recombination velocity.
Then, the adjoint problem for the flux to a given surface is obtained
by modifying only one boundary condition, namely, the one at that
surface \citep{sabelfeld17CL}.

\subsection{Spherical integral relation}

Our first goal is to derive a spherical integral equation that relates
the solution at the center of a sphere to its values on the surface
of the sphere. This equation allows a probabilistic interpretation:
the solution at the center of the sphere is equal to the mathematical
expectation over the values of the solution taken at the exit points
of the particles started at the center. This property needs to be
derived for the adjoint equations (\ref{eq:16:prime}) or (\ref{eq:16}).
We begin with the homogeneous equation (\ref{eq:16}), since the spherical
integral relation for it is easier.

The velocity $\mathbf{v}$ and the lifetime $\tau$ are assumed to
be constant in the subsequent calculations, which limits the radii
of the spheres. A corresponding diffusion length is defined as $\Lambda=\sqrt{D\tau}$.
Let us make a transformation $\tilde{w}(\mathbf{r})=\exp[(-\mathbf{v}\cdot\mathbf{r})/2D]u(\mathbf{r})$
to a new function $u(\mathbf{r})$ satisfying the following equation:
\begin{equation}
\Delta u-\lambda^{2}u=0,\label{eq:17}
\end{equation}
where $\lambda^{2}=\Lambda^{-2}+(v/2D)^{2}$ and $v=|\mathbf{v}|$.
The solution of this equation $u(\mathbf{r})$ satisfies the spherical
mean value relation \citep{sabelfeld91book}: 
\begin{equation}
u(\mathbf{r})=\frac{1}{4\pi}\frac{\lambda R}{\sinh(\lambda R)}\int u(\mathbf{r}+R\mathbf{m})d\Omega_{\mathbf{m}}.\label{eq:18}
\end{equation}
Here $R$ is the radius of a sphere with the center $\mathbf{r}$,
$\mathbf{m}$ is a unit vector from the center of the sphere, and
the integration is over the sphere of orientations of $\mathbf{m}$.
Substituting now $u(\mathbf{r})=\exp[\mathbf{v}\cdot\mathbf{r}/2D]\tilde{w}(\mathbf{r})$,
we get 
\begin{equation}
\tilde{w}(\mathbf{r})=\frac{1}{4\pi}\frac{\lambda R}{\sinh(\lambda R)}\int e^{R\mathbf{v}\cdot\mathbf{m}/2D}\tilde{w}(\mathbf{r}+R\mathbf{m})d\Omega_{\mathbf{m}}.\label{eq:19}
\end{equation}
Let us introduce a local spherical coordinate system with the $z$-axis
along the velocity $\mathbf{v}$. The zenith and the azimuthal angles
in this coordinate system are denoted by $\theta$ and $\varphi$,
so that the unit vector $\mathbf{m}$ can be written as $\mathbf{m}=(\sin\theta\cos\varphi,\sin\theta\sin\varphi,\cos\theta)$.
In this coordinate system, Eq.~(\ref{eq:19}) is given by 
\begin{equation}
\tilde{w}(\mathbf{r})=\frac{1}{4\pi}\frac{\lambda R}{\sinh(\lambda R)}\intop_{0}^{\pi}\intop_{0}^{2\pi}e^{\kappa\cos\theta}\tilde{w}(\mathbf{r}+R\mathbf{m})\sin\theta d\theta d\varphi,\label{eq:20}
\end{equation}
where $\kappa=Rv/2D$. In the laboratory frame with the $z$-axis
direction given by the surface orientation, spherical coordinates
of the velocity vector direction $\mathbf{v}/v$ are $\theta_{0}$
and $\varphi_{0}$ and the term $\exp(\kappa\cos\theta)$ in the integral
reads $\exp\left\{ \kappa\left[\sin\theta\sin\theta_{0}\cos(\varphi-\varphi_{0})+\cos\theta\cos\theta_{0}\right]\right\} .$

We now introduce the function 
\begin{equation}
p(\theta,\varphi)=\frac{\kappa\sin\theta}{4\pi\sinh\kappa}e^{\kappa\cos\theta}\label{eq:21}
\end{equation}
normalized to be a probability density, i.e., the integral of $p(\theta,\varphi)$
over $\theta$ (from $0$ to $\pi$) and over $\varphi$ (from $0$
to $2\pi$) is equal to one, and rewrite Eq.~(\ref{eq:20}) as 
\begin{equation}
\tilde{w}(\mathbf{r})=\frac{\lambda R}{\sinh(\lambda R)}\frac{\sinh\kappa}{\kappa}\intop_{0}^{\pi}\intop_{0}^{2\pi}p(\theta,\varphi)\tilde{w}(\mathbf{r}+R\mathbf{m})d\theta d\varphi.\label{eq:22}
\end{equation}
This integral relation has a clear probabilistic interpretation: a
particle starting from the center of the sphere survives with the
probability 
\begin{equation}
P_{\mathrm{surv}}(R)=\frac{\lambda R}{\sinh(\lambda R)}\frac{\sinh\kappa}{\kappa},\label{eq:23}
\end{equation}
and hits the surface of the sphere at a random exit point whose distribution
density is $p(\theta,\varphi)$.

The integral relation (\ref{eq:22}) has to be used from the “reciprocity viewpoint”: to calculate the concentration $\tilde{w}(\mathbf{r})$ at the center of a sphere, the exciton trajectory starts in the center, and gives a nonzero contribution to the concentration $\tilde{w}(\mathbf{r})$ if the exciton reaches the surface of the sphere. If the exciton annihilates, the contribution is zero. Only a portion (equal to the survival probability) of all excitons makes a nonzero contribution, so that the value $\tilde{w}(\mathbf{r})$ is smaller than the integral of $\tilde{w}(\mathbf{r})$  over the surface of the sphere. 

Next, we give an algorithm for the simulation of the exit point on
the sphere of radius $R$ centered at $\mathbf{r}$ from the probability
density (\ref{eq:21}). The probability density is axially symmetric
with respect to the direction of $\mathbf{v}$, so that the random
angles $\theta$ and $\varphi$ are independent. The angle $\varphi$
is uniformly distributed on $[0,2\pi]$, so that it is sampled as
$\varphi=2\pi\,\mathtt{rand}$, where $\mathtt{rand}$ stands for
a random number uniformly distributed on $[0,1]$. The angle $\theta$
is sampled from the density $2\pi p(\theta,\varphi)$.

To construct the generating formula for $\theta$, we introduce a
new random variable $\xi=1-\cos\theta$. A simple evaluation shows
that the distribution density of $\xi$ has the form 
\begin{equation}
f_{\xi}(x)=\frac{\kappa}{1-\exp(-2\kappa)}\exp(-\kappa x),\,\,\,\,\,0\leq x\leq2,\label{eq:24}
\end{equation}
and $f_{\xi}(x)=0$ for $x>2$. Thus, $f_{\xi}(x)$ has an exponential
distribution truncated at $x=2$. From this we find the desired simulation
formula $\cos\theta=1-\xi$, where 
\begin{equation}
\xi=-\frac{1}{\kappa}\log\left[1-(1-e^{-2\kappa})\mathtt{rand}\right].\label{eq:25}
\end{equation}
Finally, the simulation formula reads 
\begin{equation}
\cos\theta=1+\frac{1}{\kappa}\log\left[1-(1-e^{-2\kappa})\mathtt{rand}\right].\label{eq:26}
\end{equation}
Recall that the exit probability is calculated in the local coordinate
system given by the direction of $\mathbf{v}$. After the exit point
to the sphere is sampled, it is required to rotate the coordinate
system back to the reference laboratory frame and recalculate the
coordinates of the exit point accordingly. Let us give here explicit
formulae. Assume the center of the sphere in a fixed coordinate system
has coordinates $x',y',z'$, and the unit direction vector of the
velocity has coordinates $(a',b',c')$, i.e., $a'=v_{x}/|\mathbf{v}|,b'=v_{y}/|\mathbf{v}|,c'=v_{z}/|\mathbf{v}|$.
Let $\mu=\cos\theta$ be the sampled value of $\cos\theta$. Then
the coordinates of the randomly sampled exit point on the sphere are
given by $x=x'+aR,y=y'+bR,z=z'+cR$, where 
\begin{eqnarray}
a & = & a'\mu-(b'\sin\varphi+a'c'\cos\varphi)[(1-\mu^{2})/(1-c'^{2}]^{1/2}~,\nonumber \\
b & = & b'\mu+(a'\sin\varphi-b'c'\cos\varphi)[(1-\mu^{2})/(1-c'^{2}]^{1/2}~,\nonumber \\
c & = & c'\mu+(1-c'^{2})\cos\varphi[(1-\mu^{2})/(1-c'^{2}]^{1/2}~,\label{rotation}
\end{eqnarray}
and $\varphi$ is isotropic, hence $\varphi=2\pi\,\mathtt{rand}$.

Now we consider a generalization of the spherical mean value relation
to the inhomogeneous equation (\ref{eq:16:prime}). The same conditions
of constant velocity $\mathbf{v}$ and time $\tau$ are assumed within
a considered sphere. \textit{\emph{The solution $w(\mathbf{x})$ of
the drift-diffusion equation (\ref{eq:16:prime}) satisfies the following
spherical integral relation for any sphere with center $\mathbf{r}$
and radius $R$ (the sphere is entirely in the sample): 
\begin{eqnarray}
&&
w(\mathbf{r})  =  P_{\mathrm{surv}}\,\int\limits _{0}^{\pi}\int\limits _{0}^{2\pi}w(\mathbf{r}+R\mathbf{m})\,p(\theta,\varphi)\,d\varphi\,d\theta\label{ddr2}\\
 && +\, Q \,\int\limits _{0}^{2\pi}d\varphi_1\int\limits _{0}^{\pi}d\theta_1\int\limits _{0}^{R}\frac{\tau}{\tau_{r}}\,
w(\mathbf{r}+\rho\mathbf{m_1})\,
g(\rho)p_{1}(\theta_1,\varphi_1|\,\rho))\,d\rho,\nonumber 
\end{eqnarray}
where $P_{\mathrm{surv}}$ is the survival probability given by (\ref{eq:23}), and $Q=1-P_{\mathrm{surv}}$.
This relation has a clear probabilistic interpretation: with probability
$P_{\mathrm{surv}}$, the particle started in the center of the sphere
reaches its surface. The position of the particle has a distribution
on this sphere given by Eq.~(\ref{eq:21}). With probability $1-P_{\mathrm{surv}}$,
the particle does not reach the surface of the sphere, and is distributed
inside the sphere with the probability density $g(\rho)p_{1}(\theta_1,\varphi_1|\,\rho)$
where 
\begin{eqnarray}
g(\rho) & = & \frac{\rho }{(1-P_{\mathrm{surv}})D\tau}
\frac{\sinh[\mu (R-\rho)]}{\sinh(\mu R)}\,\frac{\sinh(\kappa\rho)}{\kappa},\nonumber \\
 &  & \hspace{4cm}0\le\rho\le R,\label{ddr6}
\end{eqnarray}
and 
\begin{equation}
p_{1}(\theta_1,\varphi_1|\,\rho)=\frac{\sin\theta_1}{4\pi}\,\exp(\kappa\rho\,\cos\theta_1)\,\frac{\kappa\rho}{\sinh(\kappa\rho)}~.\label{ddr7}
\end{equation}
}}To find the random point inside the sphere, one first samples a
random radius $\rho$ according to the density (\ref{ddr6}), and
then samples a random point on the sphere of radius $\rho $ according
to the angle distribution (\ref{ddr7}).

\subsection{Boundary conditions}

\label{subsec:boundary}

To complete the description of the particle trajectory simulation,
it remains to describe the interaction with the boundary. For this
purpose, we need to model the boundary conditions for the adjoint
problems. Let us write the boundary conditions (\ref{eq:12:prime})
or (\ref{eq:12}) using a finite-difference approximation of the normal
derivative in the boundary conditions at a small distance $h$ from
the boundary, with the accuracy $O(h^{2})$:

\begin{equation}
h^{-1}D\left[w(\mathbf{r})-w(\mathbf{r}-h\boldsymbol{\nu}_{k})\right]+Sw(\mathbf{r})=S\delta.\label{eq:27}
\end{equation}
Here $\delta=1/\tau_0$ in the case of boundary condition (\ref{eq:12}),
and $\delta=0$ for (\ref{eq:12:prime}). We define 
\begin{equation}
p_{h}=\frac{hS}{D+hS}\label{eq:28}
\end{equation}
and rewrite Eq.~(\ref{eq:27}) as 
\begin{equation}
w(\mathbf{r})=(1-p_{h})w(\mathbf{r}-h\boldsymbol{\nu})+p_{h}\delta.\label{eq:29}
\end{equation}
This relation can be interpreted as a total probability formula: with
probability $1-p_{h}$, a random walk trajectory is `reflected' in
the direction opposite to the normal direction $\boldsymbol{\nu}$
to the point $\mathbf{r}-h\boldsymbol{\nu}$ (assuming that this point
lies inside the domain). With the complimentary probability $p_{h}$,
the trajectory is stopped, and the solution estimate is $\delta$
[equal to $1/\tau_0$ in the case of boundary condition (\ref{eq:12}), and
$0$ for (\ref{eq:12:prime})]. The mean number of reflections is equal
to the inverse of the termination probability $p_{h}$. Therefore,
for small $h$ this number is $O(h^{-1})$, and it follows from Eq.~(\ref{eq:29})
that the total bias of the solution estimate is $O(h)$.

\subsection{Simulation of the initial distribution of excitons}

The initial distribution of excitons produced by the electron beam
$Q(\mathbf{r})$ is often approximated by the distribution of the
energy loss of the electron beam, calculated by a Monte Carlo approach
such as implemented in the free software $\mathtt{CASINO}$ \citep{drouin07}.
However, in a separate study performed on planar structures \citep{jahn18},
we found that the spatial distribution of excitons $Q(\mathbf{r})$
produced by the electron beam in GaN is notably broader, and can effectively
be described by a convolution of the energy loss distribution $Q_{0}(\mathbf{r})$
provided by $\mathtt{CASINO}$ with a Gaussian function,
\begin{equation}
Q(x,y,z)=\iintop_{-\infty}^{\infty}p(x-\xi,y-\eta)Q_{0}(\xi,\eta,z)\,d\xi\,d\eta.\label{eq:30}
\end{equation}
Here the function 
\begin{equation}
p(x,y)=(2\pi\sigma^{2})^{-1}\exp\left[-(x^{2}+y^{2})/2\sigma^{2}\right]\label{eq:31}
\end{equation}
 is normalized to be a probability density. The energy loss distribution
$Q_{\text{0}}(\mathbf{r})$ is provided by the free software $\mathtt{CASINO}$
\cite{drouin07}, and we assume that it is also normalized to be
a probability density.

Then, the composition method for sampling random numbers from the
density $Q(x,y,z)$ is as follows \citep{sabelfeld12}.

1. Generate random numbers $(\xi,\eta,z)$ from the density $Q_{0}(\xi,\eta,z)$.
We use Walker's alias algorithm \citep{walker77,devroye86} which
is extremely efficient since the number of operations it needs is
independent of the number of points.

2. With these values $(\xi,\eta)$, generate Gaussian random variables
$(x,y)$ with the mean values $(\xi,\eta)$ and the standard deviation
$\sigma$.

\subsection{Simulation of the drift-diffusion process}

\label{subsec:algorithm}

We are now in a position to describe the algorithm for the calculation
of the flux to the surface $I_{s}(\mathbf{r})$ and the CL intensity
$I_{\mathrm{CL}}(\mathbf{r})$. The Monte Carlo simulation algorithm
is based on the reciprocity theorems presented above. Note that, in
the case of pure diffusion, the situation is simpler since this equation
is self-adjoint. In our context, this fact implies that the direct
trajectories starting from the source and hitting the boundary are
statistically equivalent to the backward trajectories which start
on the boundary and come back to the source. The drift-diffusion equation,
however, is not self-adjoint and the scheme of the Monte Carlo algorithm
is more sophisticated: first, due to the reciprocity theorems we turn
to the adjoint equation, and the sign of the velocity is changed.
To solve the adjoint equation, we use the spherical integral relations,
which in turn are of `backward nature', since they relate the solution
in the starting point with the solution inside the inscribed ball
and on its boundary.

We next provide step-by-step description of the
random walk on spheres algorithm 
for solving the drift-diffusion equation.

\noindent 1. Choose the starting point of the trajectory $\mathbf{r}_{0}$,
as described in the previous section.

\noindent 2. Simulate a trajectory of an exciton starting from the
point $\mathbf{r}$:

\noindent 2.1 Choose a sphere with a radius $R$ such that the drift
velocity $\mathbf{v}$ and the time $\tau$ can be approximated as
a constant within the sphere. Practically, we check if the radii were
small enough by probe calculations until the solution is no longer
changed.

\noindent 2.2. Calculate the survival probability $P_{\mathrm{surv}}(\mathbf{r})$
given by Eq.~(\ref{eq:23}), and check if the exciton survives. If
this is the case, model the exit point on the sphere by Eqs.~(\ref{eq:26})
and (\ref{rotation}), put the exciton there, choose analogously the
second sphere of the random walk on spheres process centered at this
exit point, evaluate the survival probability, etc.

\noindent 2.3. If the exciton did not survive, add a value
of $\tau(\mathbf{r})/\tau_{r}$ for the CL intensity, as it follows
from the second term in Eq.~(\ref{ddr2}). Use the local band gap
$E_{G}(\mathbf{r})$ to obtain the energy of the emitted photon, and
add a value $E_{G}(\mathbf{r})\tau(\mathbf{r})/\tau_{r}$ to the average
energy. Then, turn to the simulation of the next trajectory starting
from the new seed point $\mathbf{r}_{0}$, i.e., go to 1.

\noindent 3. Boundary conditions.  First, define a shell $\Gamma_\varepsilon$ along the boundary $\Gamma$, such that the distance from each point in $\Gamma_\varepsilon$ to the boundary $\Gamma$ is smaller than $\varepsilon$,  which is a small distance depending on the desired accuracy. Typically, we take $\varepsilon$ equal to $10^{-3}$ of the diffusion length. If the trajectory survives in all spheres, and reaches $\Gamma_\varepsilon$, evaluate the flux:

\noindent 3.1. In the case of Dirichlet (absorption) boundary conditions,
$S\rightarrow\infty$, add a value of 1 for the intensity $I_{s}$.

\noindent 3.2. In the case of pure reflection conditions ($S=0)$,
reflect the exciton as described in Sec.~\ref{subsec:boundary},
without changing the score for the CL intensity.

\noindent 3.3. In the general case of finite surface recombination
velocity, simulate a partial reflection, i.e., with the probability
$1-p_{h}$ the exciton is reflected without adding to the CL intensity,
and with probability $p_{h}$ the trajectory stops.

\noindent 4. After simulating $N$ trajectories, the results are obtained
by averaging the total scores over all trajectories.

The accuracy of the calculation $\delta_{\varepsilon}$ is proportional
to the chosen distance $\varepsilon$ to the surface, and the reflection
step $h$ is chosen proportional to $\sqrt{\varepsilon}$. The number
of simulated trajectories $N$ is related to the desired accuracy
as $\delta_{\varepsilon}\sim N^{-1/2}$.

In step 2.3 of the algorithm, we took the center of the current sphere
$\mathbf{r}_{0}$ as a point of the recombination event, and calculated
the energy of the photon emitted in the recombination event from the
local band gap $E_{G}(\mathbf{r}_{0})$. The use of the center of
the sphere, rather than an appropriately modeled random point inside
the sphere, is sufficiently accurate in the present case, since the
variable drift velocity restricts us to the use of small spheres in
any case. We have calculated the probability density distribution
of the recombination events inside the sphere and, using it, modeled
the recombination points with higher accuracy. Such calculations are
time consuming, and we used them only to check that the use of the
center of the sphere is indeed sufficiently accurate.


\begin{thebibliography}{45}%
\makeatletter
\providecommand \@ifxundefined [1]{%
 \@ifx{#1\undefined}
}%
\providecommand \@ifnum [1]{%
 \ifnum #1\expandafter \@firstoftwo
 \else \expandafter \@secondoftwo
 \fi
}%
\providecommand \@ifx [1]{%
 \ifx #1\expandafter \@firstoftwo
 \else \expandafter \@secondoftwo
 \fi
}%
\providecommand \natexlab [1]{#1}%
\providecommand \enquote  [1]{``#1''}%
\providecommand \bibnamefont  [1]{#1}%
\providecommand \bibfnamefont [1]{#1}%
\providecommand \citenamefont [1]{#1}%
\providecommand \href@noop [0]{\@secondoftwo}%
\providecommand \href [0]{\begingroup \@sanitize@url \@href}%
\providecommand \@href[1]{\@@startlink{#1}\@@href}%
\providecommand \@@href[1]{\endgroup#1\@@endlink}%
\providecommand \@sanitize@url [0]{\catcode `\\12\catcode `\$12\catcode
  `\&12\catcode `\#12\catcode `\^12\catcode `\_12\catcode `\%12\relax}%
\providecommand \@@startlink[1]{}%
\providecommand \@@endlink[0]{}%
\providecommand \url  [0]{\begingroup\@sanitize@url \@url }%
\providecommand \@url [1]{\endgroup\@href {#1}{\urlprefix }}%
\providecommand \urlprefix  [0]{URL }%
\providecommand \Eprint [0]{\href }%
\providecommand \doibase [0]{http://dx.doi.org/}%
\providecommand \selectlanguage [0]{\@gobble}%
\providecommand \bibinfo  [0]{\@secondoftwo}%
\providecommand \bibfield  [0]{\@secondoftwo}%
\providecommand \translation [1]{[#1]}%
\providecommand \BibitemOpen [0]{}%
\providecommand \bibitemStop [0]{}%
\providecommand \bibitemNoStop [0]{.\EOS\space}%
\providecommand \EOS [0]{\spacefactor3000\relax}%
\providecommand \BibitemShut  [1]{\csname bibitem#1\endcsname}%
\let\auto@bib@innerbib\@empty
\bibitem [{\citenamefont {Lax}(1978)}]{lax78}%
  \BibitemOpen
  \bibfield  {author} {\bibinfo {author} {\bibfnamefont {M.}~\bibnamefont
  {Lax}},\ }\bibfield  {title} {\enquote {\bibinfo {title} {Junction current
  and luminescence near a dislocation or a surface},}\ }\href@noop {}
  {\bibfield  {journal} {\bibinfo  {journal} {J. Appl. Phys.}\ }\textbf
  {\bibinfo {volume} {49}},\ \bibinfo {pages} {2796} (\bibinfo {year}
  {1978})}\BibitemShut {NoStop}%
\bibitem [{\citenamefont {Donolato}(1978)}]{donolato78}%
  \BibitemOpen
  \bibfield  {author} {\bibinfo {author} {\bibfnamefont {C.}~\bibnamefont
  {Donolato}},\ }\bibfield  {title} {\enquote {\bibinfo {title} {On the theory
  of {SEM} charge-collection imaging of localized defects in semiconductors},}\
  }\href@noop {} {\bibfield  {journal} {\bibinfo  {journal} {Optik}\ }\textbf
  {\bibinfo {volume} {52}},\ \bibinfo {pages} {19} (\bibinfo {year}
  {1978})}\BibitemShut {NoStop}%
\bibitem [{\citenamefont {Donolato}(1979)}]{donolato79}%
  \BibitemOpen
  \bibfield  {author} {\bibinfo {author} {\bibfnamefont {C.}~\bibnamefont
  {Donolato}},\ }\bibfield  {title} {\enquote {\bibinfo {title} {Contrast and
  resolution of {SEM} charge-collection images of dislocations},}\ }\href@noop
  {} {\bibfield  {journal} {\bibinfo  {journal} {Appl. Phys. Lett.}\ }\textbf
  {\bibinfo {volume} {34}},\ \bibinfo {pages} {80} (\bibinfo {year}
  {1979})}\BibitemShut {NoStop}%
\bibitem [{\citenamefont {Donolato}(1985)}]{donolato85}%
  \BibitemOpen
  \bibfield  {author} {\bibinfo {author} {\bibfnamefont {C.}~\bibnamefont
  {Donolato}},\ }\bibfield  {title} {\enquote {\bibinfo {title} {A reciprocity
  theorem for charge collection},}\ }\href@noop {} {\bibfield  {journal}
  {\bibinfo  {journal} {Appl. Phys. Lett.}\ }\textbf {\bibinfo {volume} {46}},\
  \bibinfo {pages} {270} (\bibinfo {year} {1985})}\BibitemShut {NoStop}%
\bibitem [{\citenamefont {Donolato}(1998)}]{donolato98}%
  \BibitemOpen
  \bibfield  {author} {\bibinfo {author} {\bibfnamefont {C.}~\bibnamefont
  {Donolato}},\ }\bibfield  {title} {\enquote {\bibinfo {title} {Modeling the
  effect of dislocations on the minority carrier diffusion length of a
  semiconductor},}\ }\href@noop {} {\bibfield  {journal} {\bibinfo  {journal}
  {J. Appl. Phys.}\ }\textbf {\bibinfo {volume} {84}},\ \bibinfo {pages} {2656}
  (\bibinfo {year} {1998})}\BibitemShut {NoStop}%
\bibitem [{\citenamefont {Jakubowicz}(1985)}]{jakubowicz85}%
  \BibitemOpen
  \bibfield  {author} {\bibinfo {author} {\bibfnamefont {A.}~\bibnamefont
  {Jakubowicz}},\ }\bibfield  {title} {\enquote {\bibinfo {title} {On the
  theory of electron-beam-induced current contrast from pointlike defects in
  semiconductors},}\ }\href@noop {} {\bibfield  {journal} {\bibinfo  {journal}
  {J. Appl. Phys.}\ }\textbf {\bibinfo {volume} {57}},\ \bibinfo {pages} {1194}
  (\bibinfo {year} {1985})}\BibitemShut {NoStop}%
\bibitem [{\citenamefont {Jakubowicz}(1986)}]{jakubowicz86}%
  \BibitemOpen
  \bibfield  {author} {\bibinfo {author} {\bibfnamefont {A.}~\bibnamefont
  {Jakubowicz}},\ }\bibfield  {title} {\enquote {\bibinfo {title} {Theory of
  cathodoluminescence contrast from localized defects in semiconductors},}\
  }\href@noop {} {\bibfield  {journal} {\bibinfo  {journal} {J. Appl. Phys.}\
  }\textbf {\bibinfo {volume} {59}},\ \bibinfo {pages} {2205} (\bibinfo {year}
  {1986})}\BibitemShut {NoStop}%
\bibitem [{\citenamefont {Pasemann}(1991)}]{pasemann91}%
  \BibitemOpen
  \bibfield  {author} {\bibinfo {author} {\bibfnamefont {L.}~\bibnamefont
  {Pasemann}},\ }\bibfield  {title} {\enquote {\bibinfo {title} {A contribution
  to the theory of beam-induced current characterization of dislocations},}\
  }\href@noop {} {\bibfield  {journal} {\bibinfo  {journal} {J. Appl. Phys.}\
  }\textbf {\bibinfo {volume} {69}},\ \bibinfo {pages} {6387} (\bibinfo {year}
  {1991})}\BibitemShut {NoStop}%
\bibitem [{\citenamefont {Rosner}\ \emph {et~al.}(1997)\citenamefont {Rosner},
  \citenamefont {Carr}, \citenamefont {Ludowise}, \citenamefont {Girolami},\
  and\ \citenamefont {Erikson}}]{rosner97}%
  \BibitemOpen
  \bibfield  {author} {\bibinfo {author} {\bibfnamefont {S.~J.}\ \bibnamefont
  {Rosner}}, \bibinfo {author} {\bibfnamefont {E.~C.}\ \bibnamefont {Carr}},
  \bibinfo {author} {\bibfnamefont {M.~J.}\ \bibnamefont {Ludowise}}, \bibinfo
  {author} {\bibfnamefont {G.}~\bibnamefont {Girolami}}, \ and\ \bibinfo
  {author} {\bibfnamefont {H.~I.}\ \bibnamefont {Erikson}},\ }\bibfield
  {title} {\enquote {\bibinfo {title} {Correlation of cathodoluminescence
  inhomogeneity with microstructural defects in epitaxial {GaN} grown by
  metalorganic chemical-vapor deposition},}\ }\href@noop {} {\bibfield
  {journal} {\bibinfo  {journal} {Appl. Phys. Lett.}\ }\textbf {\bibinfo
  {volume} {70}},\ \bibinfo {pages} {420} (\bibinfo {year} {1997})}\BibitemShut
  {NoStop}%
\bibitem [{\citenamefont {Shmidt}\ \emph {et~al.}(2002)\citenamefont {Shmidt},
  \citenamefont {Soltanovich}, \citenamefont {Usikov}, \citenamefont
  {Yakimov},\ and\ \citenamefont {Zavarin}}]{shmidt02}%
  \BibitemOpen
  \bibfield  {author} {\bibinfo {author} {\bibfnamefont {N.~M.}\ \bibnamefont
  {Shmidt}}, \bibinfo {author} {\bibfnamefont {O.~A.}\ \bibnamefont
  {Soltanovich}}, \bibinfo {author} {\bibfnamefont {A.~S.}\ \bibnamefont
  {Usikov}}, \bibinfo {author} {\bibfnamefont {E.~B.}\ \bibnamefont {Yakimov}},
  \ and\ \bibinfo {author} {\bibfnamefont {E.~E.}\ \bibnamefont {Zavarin}},\
  }\bibfield  {title} {\enquote {\bibinfo {title} {High-resolution
  electron-beam-induced-current study of the defect structure in {GaN}
  epilayers},}\ }\href@noop {} {\bibfield  {journal} {\bibinfo  {journal} {J.
  Phys.: Condens. Matter}\ }\textbf {\bibinfo {volume} {14}},\ \bibinfo {pages}
  {13285} (\bibinfo {year} {2002})}\BibitemShut {NoStop}%
\bibitem [{\citenamefont {Nakaji}\ \emph {et~al.}(2005)\citenamefont {Nakaji},
  \citenamefont {Grillo}, \citenamefont {Yamamoto},\ and\ \citenamefont
  {Mukai}}]{nakaij05}%
  \BibitemOpen
  \bibfield  {author} {\bibinfo {author} {\bibfnamefont {D.}~\bibnamefont
  {Nakaji}}, \bibinfo {author} {\bibfnamefont {V.}~\bibnamefont {Grillo}},
  \bibinfo {author} {\bibfnamefont {N.}~\bibnamefont {Yamamoto}}, \ and\
  \bibinfo {author} {\bibfnamefont {T.}~\bibnamefont {Mukai}},\ }\bibfield
  {title} {\enquote {\bibinfo {title} {Contrast analysis of dislocation images
  in {TEM}-cathodoluminescence technique},}\ }\href@noop {} {\bibfield
  {journal} {\bibinfo  {journal} {J. Electron Microscopy}\ }\textbf {\bibinfo
  {volume} {54}},\ \bibinfo {pages} {223} (\bibinfo {year} {2005})}\BibitemShut
  {NoStop}%
\bibitem [{\citenamefont {Pauc}\ \emph {et~al.}(2006)\citenamefont {Pauc},
  \citenamefont {Phillips}, \citenamefont {Aimez},\ and\ \citenamefont
  {Drouin}}]{pauc06}%
  \BibitemOpen
  \bibfield  {author} {\bibinfo {author} {\bibfnamefont {N.}~\bibnamefont
  {Pauc}}, \bibinfo {author} {\bibfnamefont {M.~R.}\ \bibnamefont {Phillips}},
  \bibinfo {author} {\bibfnamefont {V.}~\bibnamefont {Aimez}}, \ and\ \bibinfo
  {author} {\bibfnamefont {D.}~\bibnamefont {Drouin}},\ }\bibfield  {title}
  {\enquote {\bibinfo {title} {Carrier recombination near threading
  dislocations in {GaN} epilayers by low voltage cathodoluminescence},}\
  }\href@noop {} {\bibfield  {journal} {\bibinfo  {journal} {Appl. Phys.
  Lett.}\ }\textbf {\bibinfo {volume} {89}},\ \bibinfo {pages} {161905}
  (\bibinfo {year} {2006})}\BibitemShut {NoStop}%
\bibitem [{\citenamefont {Yakimov}\ \emph {et~al.}(2007)\citenamefont
  {Yakimov}, \citenamefont {Borisov},\ and\ \citenamefont
  {Zaitsev}}]{yakimov07}%
  \BibitemOpen
  \bibfield  {author} {\bibinfo {author} {\bibfnamefont {E.~B.}\ \bibnamefont
  {Yakimov}}, \bibinfo {author} {\bibfnamefont {S.~S.}\ \bibnamefont
  {Borisov}}, \ and\ \bibinfo {author} {\bibfnamefont {S.~I.}\ \bibnamefont
  {Zaitsev}},\ }\bibfield  {title} {\enquote {\bibinfo {title} {{EBIC}
  measurements of small diffusion length in semiconductor structures},}\
  }\href@noop {} {\bibfield  {journal} {\bibinfo  {journal} {Semiconductors}\
  }\textbf {\bibinfo {volume} {41}},\ \bibinfo {pages} {411} (\bibinfo {year}
  {2007})}\BibitemShut {NoStop}%
\bibitem [{\citenamefont {Ino}\ and\ \citenamefont {Yamamoto}(2008)}]{ino08}%
  \BibitemOpen
  \bibfield  {author} {\bibinfo {author} {\bibfnamefont {N.}~\bibnamefont
  {Ino}}\ and\ \bibinfo {author} {\bibfnamefont {N.}~\bibnamefont {Yamamoto}},\
  }\bibfield  {title} {\enquote {\bibinfo {title} {Low temperature diffusion
  length of excitons in gallium nitride measured by cathodoluminescence
  technique},}\ }\href@noop {} {\bibfield  {journal} {\bibinfo  {journal}
  {Appl. Phys. Lett.}\ }\textbf {\bibinfo {volume} {93}},\ \bibinfo {pages}
  {232103} (\bibinfo {year} {2008})}\BibitemShut {NoStop}%
\bibitem [{\citenamefont {Yakimov}(2010)}]{yakimov10}%
  \BibitemOpen
  \bibfield  {author} {\bibinfo {author} {\bibfnamefont {E.~B.}\ \bibnamefont
  {Yakimov}},\ }\bibfield  {title} {\enquote {\bibinfo {title} {{Comment on
  ``Carrier recombination near threading dislocations in {GaN} epilayers by low
  voltage cathodoluminescence [Appl. Phys. Lett. 89, 161905, 2006]"}},}\
  }\href@noop {} {\bibfield  {journal} {\bibinfo  {journal} {Appl. Phys.
  Lett.}\ }\textbf {\bibinfo {volume} {97}},\ \bibinfo {pages} {166101}
  (\bibinfo {year} {2010})}\BibitemShut {NoStop}%
\bibitem [{\citenamefont {Yakimov}(2015)}]{yakimov15}%
  \BibitemOpen
  \bibfield  {author} {\bibinfo {author} {\bibfnamefont {E.~B.}\ \bibnamefont
  {Yakimov}},\ }\bibfield  {title} {\enquote {\bibinfo {title} {What is the
  real value of diffusion length in {GaN}?}}\ }\href@noop {} {\bibfield
  {journal} {\bibinfo  {journal} {J. Alloys Compounds}\ }\textbf {\bibinfo
  {volume} {627}},\ \bibinfo {pages} {344} (\bibinfo {year}
  {2015})}\BibitemShut {NoStop}%
\bibitem [{\citenamefont {Sabelfeld}\ \emph {et~al.}(2017)\citenamefont
  {Sabelfeld}, \citenamefont {Kaganer}, \citenamefont {Pf\"{u}ller},\ and\
  \citenamefont {Brandt}}]{sabelfeld17CL}%
  \BibitemOpen
  \bibfield  {author} {\bibinfo {author} {\bibfnamefont {K.~K.}\ \bibnamefont
  {Sabelfeld}}, \bibinfo {author} {\bibfnamefont {V.~M.}\ \bibnamefont
  {Kaganer}}, \bibinfo {author} {\bibfnamefont {C.}~\bibnamefont
  {Pf\"{u}ller}}, \ and\ \bibinfo {author} {\bibfnamefont {O.}~\bibnamefont
  {Brandt}},\ }\bibfield  {title} {\enquote {\bibinfo {title} {Dislocation
  contrast in cathodoluminescence and electron-beam induced current maps on
  {GaN(0001)}},}\ }\href@noop {} {\bibfield  {journal} {\bibinfo  {journal} {J.
  Phys. D: Appl. Phys.}\ }\textbf {\bibinfo {volume} {50}},\ \bibinfo {pages}
  {405101} (\bibinfo {year} {2017})}\BibitemShut {NoStop}%
\bibitem [{\citenamefont {Kaganer}\ \emph {et~al.}(2018)\citenamefont
  {Kaganer}, \citenamefont {Sabelfeld},\ and\ \citenamefont
  {Brandt}}]{kaganer18apl}%
  \BibitemOpen
  \bibfield  {author} {\bibinfo {author} {\bibfnamefont {V.~M.}\ \bibnamefont
  {Kaganer}}, \bibinfo {author} {\bibfnamefont {K.~K.}\ \bibnamefont
  {Sabelfeld}}, \ and\ \bibinfo {author} {\bibfnamefont {O.}~\bibnamefont
  {Brandt}},\ }\bibfield  {title} {\enquote {\bibinfo {title} {Piezoelectric
  field, exciton lifetime, and cathodoluminescence intensity at threading
  dislocations in {GaN\{0001\}}},}\ }\href@noop {} {\bibfield  {journal}
  {\bibinfo  {journal} {Appl. Phys. Lett.}\ }\textbf {\bibinfo {volume}
  {112}},\ \bibinfo {pages} {122101} (\bibinfo {year} {2018})}\BibitemShut
  {NoStop}%
\bibitem [{\citenamefont {Hirth}\ and\ \citenamefont
  {Lothe}(1982)}]{hirthlothe82}%
  \BibitemOpen
  \bibfield  {author} {\bibinfo {author} {\bibfnamefont {J.~P.}\ \bibnamefont
  {Hirth}}\ and\ \bibinfo {author} {\bibfnamefont {J.}~\bibnamefont {Lothe}},\
  }\href@noop {} {\emph {\bibinfo {title} {Theory of Dislocations}}}\ (\bibinfo
   {publisher} {Wiley},\ \bibinfo {address} {N. Y.},\ \bibinfo {year}
  {1982})\BibitemShut {NoStop}%
\bibitem [{\citenamefont {Indenbom}\ and\ \citenamefont
  {Lothe}(1992)}]{IndenbomLothe}%
  \BibitemOpen
  \bibinfo {editor} {\bibfnamefont {V.~L.}\ \bibnamefont {Indenbom}}\ and\
  \bibinfo {editor} {\bibfnamefont {J.}~\bibnamefont {Lothe}},\ eds.,\
  \href@noop {} {\emph {\bibinfo {title} {Elastic Strain Fields and Dislocation
  Mobility}}}\ (\bibinfo  {publisher} {North-Holland},\ \bibinfo {address}
  {Amsterdam},\ \bibinfo {year} {1992})\BibitemShut {NoStop}%
\bibitem [{\citenamefont {Bir}\ and\ \citenamefont {Pikus}(1974)}]{BirPikus74}%
  \BibitemOpen
  \bibfield  {author} {\bibinfo {author} {\bibfnamefont {G.~L.}\ \bibnamefont
  {Bir}}\ and\ \bibinfo {author} {\bibfnamefont {G.~E.}\ \bibnamefont
  {Pikus}},\ }\href@noop {} {\emph {\bibinfo {title} {Symmetry and Strain
  Induced Effects in Semicounductors}}}\ (\bibinfo  {publisher} {Wiley},\
  \bibinfo {address} {New York},\ \bibinfo {year} {1974})\BibitemShut {NoStop}%
\bibitem [{\citenamefont {Ghosh}\ \emph {et~al.}(2002)\citenamefont {Ghosh},
  \citenamefont {Waltereit}, \citenamefont {Brandt}, \citenamefont {Grahn},\
  and\ \citenamefont {Ploog}}]{ghosh02}%
  \BibitemOpen
  \bibfield  {author} {\bibinfo {author} {\bibfnamefont {S.}~\bibnamefont
  {Ghosh}}, \bibinfo {author} {\bibfnamefont {P.}~\bibnamefont {Waltereit}},
  \bibinfo {author} {\bibfnamefont {O.}~\bibnamefont {Brandt}}, \bibinfo
  {author} {\bibfnamefont {H.~T.}\ \bibnamefont {Grahn}}, \ and\ \bibinfo
  {author} {\bibfnamefont {K.~H.}\ \bibnamefont {Ploog}},\ }\bibfield  {title}
  {\enquote {\bibinfo {title} {Electronic band structure of wurtzite {GaN}
  under biaxial strain in the {M} plane investigated with photoreflectance
  spectroscopy},}\ }\href@noop {} {\bibfield  {journal} {\bibinfo  {journal}
  {Phys. Rev. B}\ }\textbf {\bibinfo {volume} {65}},\ \bibinfo {pages} {075202}
  (\bibinfo {year} {2002})}\BibitemShut {NoStop}%
\bibitem [{\citenamefont {Yan}\ \emph {et~al.}(2009)\citenamefont {Yan},
  \citenamefont {Rinke}, \citenamefont {Scheffler},\ and\ \citenamefont {{Van
  de Walle}}}]{yan09}%
  \BibitemOpen
  \bibfield  {author} {\bibinfo {author} {\bibfnamefont {Q.}~\bibnamefont
  {Yan}}, \bibinfo {author} {\bibfnamefont {P.}~\bibnamefont {Rinke}}, \bibinfo
  {author} {\bibfnamefont {M.}~\bibnamefont {Scheffler}}, \ and\ \bibinfo
  {author} {\bibfnamefont {C.~G.}\ \bibnamefont {{Van de Walle}}},\ }\bibfield
  {title} {\enquote {\bibinfo {title} {Strain effects in group-iii nitrides:
  Deformation potentials for {AlN, GaN, and InN}},}\ }\href@noop {} {\bibfield
  {journal} {\bibinfo  {journal} {Appl. Phys. Lett.}\ }\textbf {\bibinfo
  {volume} {95}},\ \bibinfo {pages} {121111} (\bibinfo {year}
  {2009})}\BibitemShut {NoStop}%
\bibitem [{\citenamefont {Ishii}\ \emph {et~al.}(2010)\citenamefont {Ishii},
  \citenamefont {Kaneta}, \citenamefont {Funato}, \citenamefont {Kawakami},\
  and\ \citenamefont {Yamaguchi}}]{ishii10}%
  \BibitemOpen
  \bibfield  {author} {\bibinfo {author} {\bibfnamefont {R.}~\bibnamefont
  {Ishii}}, \bibinfo {author} {\bibfnamefont {A.}~\bibnamefont {Kaneta}},
  \bibinfo {author} {\bibfnamefont {M.}~\bibnamefont {Funato}}, \bibinfo
  {author} {\bibfnamefont {Y.}~\bibnamefont {Kawakami}}, \ and\ \bibinfo
  {author} {\bibfnamefont {A.~A.}\ \bibnamefont {Yamaguchi}},\ }\bibfield
  {title} {\enquote {\bibinfo {title} {All deformation potentials in {GaN}
  determined by reflectance spectroscopy under uniaxial stress: Definite
  breakdown of the quasicubic approximation},}\ }\href {\doibase
  10.1103/PhysRevB.81.155202} {\bibfield  {journal} {\bibinfo  {journal} {Phys.
  Rev. B}\ }\textbf {\bibinfo {volume} {81}},\ \bibinfo {pages} {155202}
  (\bibinfo {year} {2010})}\BibitemShut {NoStop}%
\bibitem [{\citenamefont {Sabelfeld}(1991)}]{sabelfeld91book}%
  \BibitemOpen
  \bibfield  {author} {\bibinfo {author} {\bibfnamefont {K.~K.}\ \bibnamefont
  {Sabelfeld}},\ }\href@noop {} {\emph {\bibinfo {title} {Monte Carlo Methods
  in Boundary Value Problems}}}\ (\bibinfo  {publisher} {Springer},\ \bibinfo
  {address} {N. Y.},\ \bibinfo {year} {1991})\BibitemShut {NoStop}%
\bibitem [{\citenamefont {Sabelfeld}(2016)}]{sabelfeld16}%
  \BibitemOpen
  \bibfield  {author} {\bibinfo {author} {\bibfnamefont {K.~K.}\ \bibnamefont
  {Sabelfeld}},\ }\bibfield  {title} {\enquote {\bibinfo {title} {Random walk
  on spheres method fo solving drift-diffusion problems},}\ }\href@noop {}
  {\bibfield  {journal} {\bibinfo  {journal} {Monte Carlo Methods Appl.}\
  }\textbf {\bibinfo {volume} {22}},\ \bibinfo {pages} {265--275} (\bibinfo
  {year} {2016})}\BibitemShut {NoStop}%
\bibitem [{\citenamefont {{de la Pe\~na}}\ \emph {et~al.}()\citenamefont {{de
  la Pe\~na}}, \citenamefont {Fauske}, \citenamefont {Burdet}, \citenamefont
  {Prestat}, \citenamefont {Jokubauskas}, \citenamefont {Nord}, \citenamefont
  {Ostasevicius}, \citenamefont {MacArthur}, \citenamefont {Sarahan},
  \citenamefont {Johnstone}, \citenamefont {Taillon}, \citenamefont {Eljarrat},
  \citenamefont {Migunov}, \citenamefont {Caron}, \citenamefont {Furnival},
  \citenamefont {Mazzucco}, \citenamefont {Aarholt}, \citenamefont {Walls},
  \citenamefont {Slater}, \citenamefont {Winkler}, \citenamefont {Martineau},
  \citenamefont {Donval}, \citenamefont {McLeod}, \citenamefont {Hoglund},
  \citenamefont {Alxneit}, \citenamefont {Hjorth}, \citenamefont {Henninen},
  \citenamefont {Zagonel}, \citenamefont {Garmannslund},\ and\ \citenamefont
  {Skorikov}}]{hiperspy}%
  \BibitemOpen
  \bibfield  {author} {\bibinfo {author} {\bibfnamefont {F.}~\bibnamefont {{de
  la Pe\~na}}}, \bibinfo {author} {\bibfnamefont {V.~T.}\ \bibnamefont
  {Fauske}}, \bibinfo {author} {\bibfnamefont {P.}~\bibnamefont {Burdet}},
  \bibinfo {author} {\bibfnamefont {E.}~\bibnamefont {Prestat}}, \bibinfo
  {author} {\bibfnamefont {P.}~\bibnamefont {Jokubauskas}}, \bibinfo {author}
  {\bibfnamefont {M.}~\bibnamefont {Nord}}, \bibinfo {author} {\bibfnamefont
  {T.}~\bibnamefont {Ostasevicius}}, \bibinfo {author} {\bibfnamefont {K.~E.}\
  \bibnamefont {MacArthur}}, \bibinfo {author} {\bibfnamefont {M.}~\bibnamefont
  {Sarahan}}, \bibinfo {author} {\bibfnamefont {D.~N.}\ \bibnamefont
  {Johnstone}}, \bibinfo {author} {\bibfnamefont {J.}~\bibnamefont {Taillon}},
  \bibinfo {author} {\bibfnamefont {A.}~\bibnamefont {Eljarrat}}, \bibinfo
  {author} {\bibfnamefont {V.}~\bibnamefont {Migunov}}, \bibinfo {author}
  {\bibfnamefont {J.}~\bibnamefont {Caron}}, \bibinfo {author} {\bibfnamefont
  {T.}~\bibnamefont {Furnival}}, \bibinfo {author} {\bibfnamefont
  {S.}~\bibnamefont {Mazzucco}}, \bibinfo {author} {\bibfnamefont
  {T.}~\bibnamefont {Aarholt}}, \bibinfo {author} {\bibfnamefont
  {M.}~\bibnamefont {Walls}}, \bibinfo {author} {\bibfnamefont
  {T.}~\bibnamefont {Slater}}, \bibinfo {author} {\bibfnamefont
  {F.}~\bibnamefont {Winkler}}, \bibinfo {author} {\bibfnamefont
  {B.}~\bibnamefont {Martineau}}, \bibinfo {author} {\bibfnamefont
  {G.}~\bibnamefont {Donval}}, \bibinfo {author} {\bibfnamefont
  {R.}~\bibnamefont {McLeod}}, \bibinfo {author} {\bibfnamefont {E.~R.}\
  \bibnamefont {Hoglund}}, \bibinfo {author} {\bibfnamefont {I.}~\bibnamefont
  {Alxneit}}, \bibinfo {author} {\bibfnamefont {I.}~\bibnamefont {Hjorth}},
  \bibinfo {author} {\bibfnamefont {T.}~\bibnamefont {Henninen}}, \bibinfo
  {author} {\bibfnamefont {L.~F.}\ \bibnamefont {Zagonel}}, \bibinfo {author}
  {\bibfnamefont {A.}~\bibnamefont {Garmannslund}}, \ and\ \bibinfo {author}
  {\bibfnamefont {A.}~\bibnamefont {Skorikov}},\ }\href@noop {} {\enquote
  {\bibinfo {title} {{HyperSpy}},}\ }\bibinfo {note} {DOI:
  10.5281/zenodo.1469364}\BibitemShut {NoStop}%
\bibitem [{\citenamefont {Aleksiej\={u}nas}\ \emph {et~al.}(2003)\citenamefont
  {Aleksiej\={u}nas}, \citenamefont {S\=ud\v{z}ius}, \citenamefont
  {Malinauskas}, \citenamefont {Vaitkus}, \citenamefont {Jara\v{s}i\={u}nas},\
  and\ \citenamefont {Sakai}}]{Aleksiejunas2003}%
  \BibitemOpen
  \bibfield  {author} {\bibinfo {author} {\bibfnamefont {R.}~\bibnamefont
  {Aleksiej\={u}nas}}, \bibinfo {author} {\bibfnamefont {M.}~\bibnamefont
  {S\=ud\v{z}ius}}, \bibinfo {author} {\bibfnamefont {T.}~\bibnamefont
  {Malinauskas}}, \bibinfo {author} {\bibfnamefont {J.}~\bibnamefont
  {Vaitkus}}, \bibinfo {author} {\bibfnamefont {K.}~\bibnamefont
  {Jara\v{s}i\={u}nas}}, \ and\ \bibinfo {author} {\bibfnamefont
  {S.}~\bibnamefont {Sakai}},\ }\bibfield  {title} {\enquote {\bibinfo {title}
  {Determination of free carrier bipolar diffusion coefficient and surface
  recombination velocity of undoped {GaN} epilayers},}\ }\href@noop {}
  {\bibfield  {journal} {\bibinfo  {journal} {Appl. Phys. Lett.}\ }\textbf
  {\bibinfo {volume} {83}},\ \bibinfo {pages} {1157} (\bibinfo {year}
  {2003})}\BibitemShut {NoStop}%
\bibitem [{\citenamefont {\v{S}\v{c}ajev}\ \emph {et~al.}(2012)\citenamefont
  {\v{S}\v{c}ajev}, \citenamefont {Jara\v{s}i\={u}nas}, \citenamefont {Okur},
  \citenamefont {\"{O}zg\"{u}r},\ and\ \citenamefont {Morko\c{c}}}]{scajev12}%
  \BibitemOpen
  \bibfield  {author} {\bibinfo {author} {\bibfnamefont {P.}~\bibnamefont
  {\v{S}\v{c}ajev}}, \bibinfo {author} {\bibfnamefont {K.}~\bibnamefont
  {Jara\v{s}i\={u}nas}}, \bibinfo {author} {\bibfnamefont {S.}~\bibnamefont
  {Okur}}, \bibinfo {author} {\bibfnamefont {\"{U}.}\ \bibnamefont
  {\"{O}zg\"{u}r}}, \ and\ \bibinfo {author} {\bibfnamefont {H.}~\bibnamefont
  {Morko\c{c}}},\ }\bibfield  {title} {\enquote {\bibinfo {title} {Carrier
  dynamics in bulk {GaN}},}\ }\href@noop {} {\bibfield  {journal} {\bibinfo
  {journal} {J. Appl. Phys.}\ }\textbf {\bibinfo {volume} {111}},\ \bibinfo
  {pages} {023702} (\bibinfo {year} {2012})}\BibitemShut {NoStop}%
\bibitem [{\citenamefont {Ebeling}\ \emph {et~al.}(1976)\citenamefont
  {Ebeling}, \citenamefont {Kraeft},\ and\ \citenamefont {Kremp}}]{ebeling76}%
  \BibitemOpen
  \bibfield  {author} {\bibinfo {author} {\bibfnamefont {W.}~\bibnamefont
  {Ebeling}}, \bibinfo {author} {\bibfnamefont {W.-D.}\ \bibnamefont {Kraeft}},
  \ and\ \bibinfo {author} {\bibfnamefont {D.}~\bibnamefont {Kremp}},\
  }\href@noop {} {\emph {\bibinfo {title} {Theory of Bound States and
  Ionization Equilibrium in Plasmas and Solids}}}\ (\bibinfo  {publisher}
  {Akademie-Verlag},\ \bibinfo {address} {Berlin},\ \bibinfo {year}
  {1976})\BibitemShut {NoStop}%
\bibitem [{\citenamefont {Gmeinwieser}\ \emph {et~al.}(2005)\citenamefont
  {Gmeinwieser}, \citenamefont {Gottfriedsen}, \citenamefont {Schwarz},
  \citenamefont {Wegscheider}, \citenamefont {Clos}, \citenamefont {Krtschil},
  \citenamefont {Krost}, \citenamefont {Weimar}, \citenamefont {Brüderl},
  \citenamefont {Lell},\ and\ \citenamefont {Härle}}]{gmeinwieser05}%
  \BibitemOpen
  \bibfield  {author} {\bibinfo {author} {\bibfnamefont {N.}~\bibnamefont
  {Gmeinwieser}}, \bibinfo {author} {\bibfnamefont {P.}~\bibnamefont
  {Gottfriedsen}}, \bibinfo {author} {\bibfnamefont {U.~T.}\ \bibnamefont
  {Schwarz}}, \bibinfo {author} {\bibfnamefont {W.}~\bibnamefont
  {Wegscheider}}, \bibinfo {author} {\bibfnamefont {R.}~\bibnamefont {Clos}},
  \bibinfo {author} {\bibfnamefont {A.}~\bibnamefont {Krtschil}}, \bibinfo
  {author} {\bibfnamefont {A.}~\bibnamefont {Krost}}, \bibinfo {author}
  {\bibfnamefont {A.}~\bibnamefont {Weimar}}, \bibinfo {author} {\bibfnamefont
  {G.}~\bibnamefont {Brüderl}}, \bibinfo {author} {\bibfnamefont
  {A.}~\bibnamefont {Lell}}, \ and\ \bibinfo {author} {\bibfnamefont
  {V.}~\bibnamefont {Härle}},\ }\bibfield  {title} {\enquote {\bibinfo {title}
  {Local strain and potential distribution induced by single dislocations in
  {GaN}},}\ }\href@noop {} {\bibfield  {journal} {\bibinfo  {journal} {J. Appl.
  Phys.}\ }\textbf {\bibinfo {volume} {98}},\ \bibinfo {pages} {116102}
  (\bibinfo {year} {2005})}\BibitemShut {NoStop}%
\bibitem [{\citenamefont {Liu}\ \emph {et~al.}(2016)\citenamefont {Liu},
  \citenamefont {Carlin}, \citenamefont {Grandjean}, \citenamefont {Deveaud},\
  and\ \citenamefont {Jacopin}}]{liu16}%
  \BibitemOpen
  \bibfield  {author} {\bibinfo {author} {\bibfnamefont {W.}~\bibnamefont
  {Liu}}, \bibinfo {author} {\bibfnamefont {J.-F.}\ \bibnamefont {Carlin}},
  \bibinfo {author} {\bibfnamefont {N.}~\bibnamefont {Grandjean}}, \bibinfo
  {author} {\bibfnamefont {B.}~\bibnamefont {Deveaud}}, \ and\ \bibinfo
  {author} {\bibfnamefont {G.}~\bibnamefont {Jacopin}},\ }\bibfield  {title}
  {\enquote {\bibinfo {title} {Exciton dynamics at a single dislocation in
  {GaN} probed by picosecond time-resolved cathodoluminescence},}\ }\href@noop
  {} {\bibfield  {journal} {\bibinfo  {journal} {Appl. Phys. Lett.}\ }\textbf
  {\bibinfo {volume} {109}},\ \bibinfo {pages} {042101} (\bibinfo {year}
  {2016})}\BibitemShut {NoStop}%
\bibitem [{\citenamefont {Tabet}(1998{\natexlab{a}})}]{tabet98ssp}%
  \BibitemOpen
  \bibfield  {author} {\bibinfo {author} {\bibfnamefont {N.}~\bibnamefont
  {Tabet}},\ }\bibfield  {title} {\enquote {\bibinfo {title} {{Monte Carlo}
  simulation of the recombination contrast of dislocations},}\ }\href@noop {}
  {\bibfield  {journal} {\bibinfo  {journal} {Sol. State Phenom.}\ }\textbf
  {\bibinfo {volume} {63--64}},\ \bibinfo {pages} {89--96} (\bibinfo {year}
  {1998}{\natexlab{a}})}\BibitemShut {NoStop}%
\bibitem [{\citenamefont {Tabet}(1998{\natexlab{b}})}]{tabet98sst}%
  \BibitemOpen
  \bibfield  {author} {\bibinfo {author} {\bibfnamefont {N.}~\bibnamefont
  {Tabet}},\ }\bibfield  {title} {\enquote {\bibinfo {title} {{Monte Carlo}
  simulation of the charge collection contrast of spherical defects in
  semiconductors},}\ }\href@noop {} {\bibfield  {journal} {\bibinfo  {journal}
  {Semicond. Sci. Technol.}\ }\textbf {\bibinfo {volume} {13}},\ \bibinfo
  {pages} {1392} (\bibinfo {year} {1998}{\natexlab{b}})}\BibitemShut {NoStop}%
\bibitem [{\citenamefont {Ledra}\ and\ \citenamefont {Tabet}(2005)}]{ledra05}%
  \BibitemOpen
  \bibfield  {author} {\bibinfo {author} {\bibfnamefont {M.}~\bibnamefont
  {Ledra}}\ and\ \bibinfo {author} {\bibfnamefont {N.}~\bibnamefont {Tabet}},\
  }\bibfield  {title} {\enquote {\bibinfo {title} {{Monte Carlo} simulation of
  the contrast of {SEM} charge-collection images of dislocations in
  semiconductors},}\ }\href@noop {} {\bibfield  {journal} {\bibinfo  {journal}
  {J. Phys. D: Appl. Phys.}\ }\textbf {\bibinfo {volume} {38}},\ \bibinfo
  {pages} {3845--3849} (\bibinfo {year} {2005})}\BibitemShut {NoStop}%
\bibitem [{\citenamefont {Monemar}\ \emph {et~al.}(2008)\citenamefont
  {Monemar}, \citenamefont {Paskov}, \citenamefont {Bergman}, \citenamefont
  {Toropov}, \citenamefont {Shubina}, \citenamefont {Malinauskas},\ and\
  \citenamefont {Usui}}]{monemar08}%
  \BibitemOpen
  \bibfield  {author} {\bibinfo {author} {\bibfnamefont {B.}~\bibnamefont
  {Monemar}}, \bibinfo {author} {\bibfnamefont {P.~P.}\ \bibnamefont {Paskov}},
  \bibinfo {author} {\bibfnamefont {J.~P.}\ \bibnamefont {Bergman}}, \bibinfo
  {author} {\bibfnamefont {A.~A.}\ \bibnamefont {Toropov}}, \bibinfo {author}
  {\bibfnamefont {T.~V.}\ \bibnamefont {Shubina}}, \bibinfo {author}
  {\bibfnamefont {T.}~\bibnamefont {Malinauskas}}, \ and\ \bibinfo {author}
  {\bibfnamefont {A.}~\bibnamefont {Usui}},\ }\bibfield  {title} {\enquote
  {\bibinfo {title} {Recombination of free and bound excitons in {GaN}},}\
  }\href@noop {} {\bibfield  {journal} {\bibinfo  {journal} {Phys. Stat. Sol.
  (b)}\ }\textbf {\bibinfo {volume} {245}},\ \bibinfo {pages} {1723--1740}
  (\bibinfo {year} {2008})}\BibitemShut {NoStop}%
\bibitem [{\citenamefont {{J. A. Freitas, Jr.}}\ \emph
  {et~al.}(2002)\citenamefont {{J. A. Freitas, Jr.}}, \citenamefont {Moore},
  \citenamefont {Shanabrook}, \citenamefont {Braga}, \citenamefont {Lee},
  \citenamefont {Park},\ and\ \citenamefont {Han}}]{freitas02}%
  \BibitemOpen
  \bibfield  {author} {\bibinfo {author} {\bibnamefont {{J. A. Freitas, Jr.}}},
  \bibinfo {author} {\bibfnamefont {W.~J.}\ \bibnamefont {Moore}}, \bibinfo
  {author} {\bibfnamefont {B.~V.}\ \bibnamefont {Shanabrook}}, \bibinfo
  {author} {\bibfnamefont {G.~C.~B.}\ \bibnamefont {Braga}}, \bibinfo {author}
  {\bibfnamefont {S.~K.}\ \bibnamefont {Lee}}, \bibinfo {author} {\bibfnamefont
  {S.~S.}\ \bibnamefont {Park}}, \ and\ \bibinfo {author} {\bibfnamefont
  {J.~Y.}\ \bibnamefont {Han}},\ }\bibfield  {title} {\enquote {\bibinfo
  {title} {Donor-related recombination processes in hydride-vapor-phase
  epitaxial {GaN}},}\ }\href@noop {} {\bibfield  {journal} {\bibinfo  {journal}
  {Phys. Rev. B}\ }\textbf {\bibinfo {volume} {66}},\ \bibinfo {pages} {233311}
  (\bibinfo {year} {2002})}\BibitemShut {NoStop}%
\bibitem [{\citenamefont {Wysmolek}\ \emph {et~al.}(2002)\citenamefont
  {Wysmolek}, \citenamefont {Korona}, \citenamefont {St\c{e}pniewski},
  \citenamefont {Baranowski}, \citenamefont {B{\l}oniarz}, \citenamefont
  {Potemski}, \citenamefont {Jones}, \citenamefont {Look}, \citenamefont
  {Kuhl}, \citenamefont {Park},\ and\ \citenamefont {Lee}}]{wysmolek02}%
  \BibitemOpen
  \bibfield  {author} {\bibinfo {author} {\bibfnamefont {A.}~\bibnamefont
  {Wysmolek}}, \bibinfo {author} {\bibfnamefont {K.~P.}\ \bibnamefont
  {Korona}}, \bibinfo {author} {\bibfnamefont {R.}~\bibnamefont
  {St\c{e}pniewski}}, \bibinfo {author} {\bibfnamefont {J.~M.}\ \bibnamefont
  {Baranowski}}, \bibinfo {author} {\bibfnamefont {J.}~\bibnamefont
  {B{\l}oniarz}}, \bibinfo {author} {\bibfnamefont {M.}~\bibnamefont
  {Potemski}}, \bibinfo {author} {\bibfnamefont {R.~L.}\ \bibnamefont {Jones}},
  \bibinfo {author} {\bibfnamefont {D.~C.}\ \bibnamefont {Look}}, \bibinfo
  {author} {\bibfnamefont {J.}~\bibnamefont {Kuhl}}, \bibinfo {author}
  {\bibfnamefont {S.~S.}\ \bibnamefont {Park}}, \ and\ \bibinfo {author}
  {\bibfnamefont {S.~K.}\ \bibnamefont {Lee}},\ }\bibfield  {title} {\enquote
  {\bibinfo {title} {Recombination of excitons bound to oxygen and silicon
  donors in freestanding {GaN}},}\ }\href@noop {} {\bibfield  {journal}
  {\bibinfo  {journal} {Phys. Rev. B}\ }\textbf {\bibinfo {volume} {66}},\
  \bibinfo {pages} {245317} (\bibinfo {year} {2002})}\BibitemShut {NoStop}%
\bibitem [{\citenamefont {Monemar}\ \emph {et~al.}(1996)\citenamefont
  {Monemar}, \citenamefont {Bergman}, \citenamefont {Buyanova}, \citenamefont
  {Li}, \citenamefont {Amano},\ and\ \citenamefont {Akasaki}}]{monemar96}%
  \BibitemOpen
  \bibfield  {author} {\bibinfo {author} {\bibfnamefont {B.}~\bibnamefont
  {Monemar}}, \bibinfo {author} {\bibfnamefont {J.~P.}\ \bibnamefont
  {Bergman}}, \bibinfo {author} {\bibfnamefont {I.~A.}\ \bibnamefont
  {Buyanova}}, \bibinfo {author} {\bibfnamefont {W.}~\bibnamefont {Li}},
  \bibinfo {author} {\bibfnamefont {H.}~\bibnamefont {Amano}}, \ and\ \bibinfo
  {author} {\bibfnamefont {I.}~\bibnamefont {Akasaki}},\ }\bibfield  {title}
  {\enquote {\bibinfo {title} {Free excitons in {GaN}},}\ }\href@noop {}
  {\bibfield  {journal} {\bibinfo  {journal} {MRS Internet J. Nitride Semicond.
  Res.}\ }\textbf {\bibinfo {volume} {1}},\ \bibinfo {pages} {{e2}} (\bibinfo
  {year} {1996})}\BibitemShut {NoStop}%
\bibitem [{\citenamefont {Kovalev}\ \emph {et~al.}(1996)\citenamefont
  {Kovalev}, \citenamefont {Averboukh}, \citenamefont {Volm}, \citenamefont
  {Meyer}, \citenamefont {Amano},\ and\ \citenamefont {Akasaki}}]{kovalev96}%
  \BibitemOpen
  \bibfield  {author} {\bibinfo {author} {\bibfnamefont {D.}~\bibnamefont
  {Kovalev}}, \bibinfo {author} {\bibfnamefont {B.}~\bibnamefont {Averboukh}},
  \bibinfo {author} {\bibfnamefont {D.}~\bibnamefont {Volm}}, \bibinfo {author}
  {\bibfnamefont {B.~K.}\ \bibnamefont {Meyer}}, \bibinfo {author}
  {\bibfnamefont {H.}~\bibnamefont {Amano}}, \ and\ \bibinfo {author}
  {\bibfnamefont {I.}~\bibnamefont {Akasaki}},\ }\bibfield  {title} {\enquote
  {\bibinfo {title} {Free exciton emission in {GaN}},}\ }\href@noop {}
  {\bibfield  {journal} {\bibinfo  {journal} {Phys. Rev. B}\ }\textbf {\bibinfo
  {volume} {54}},\ \bibinfo {pages} {2518--2522} (\bibinfo {year}
  {1996})}\BibitemShut {NoStop}%
\bibitem [{\citenamefont {Drouin}\ \emph {et~al.}(2007)\citenamefont {Drouin},
  \citenamefont {Couture}, \citenamefont {Joly}, \citenamefont {Tastet},\ and\
  \citenamefont {Aimez}}]{drouin07}%
  \BibitemOpen
  \bibfield  {author} {\bibinfo {author} {\bibfnamefont {D.}~\bibnamefont
  {Drouin}}, \bibinfo {author} {\bibfnamefont {A.~R.}\ \bibnamefont {Couture}},
  \bibinfo {author} {\bibfnamefont {D.}~\bibnamefont {Joly}}, \bibinfo {author}
  {\bibfnamefont {X.}~\bibnamefont {Tastet}}, \ and\ \bibinfo {author}
  {\bibfnamefont {V.}~\bibnamefont {Aimez}},\ }\bibfield  {title} {\enquote
  {\bibinfo {title} {Casino v2.42 --- a fast and easy-to-use modeling tool for
  scanning electron microscopy and microanalysis users},}\ }\href@noop {}
  {\bibfield  {journal} {\bibinfo  {journal} {Scanning}\ }\textbf {\bibinfo
  {volume} {29}},\ \bibinfo {pages} {92--101} (\bibinfo {year}
  {2007})}\BibitemShut {NoStop}%
\bibitem [{jah()}]{jahn18}%
  \BibitemOpen
  \href@noop {} {}\bibinfo {note} {U. Jahn et al., to be published}\BibitemShut
  {NoStop}%
\bibitem [{\citenamefont {Sabelfeld}(2012)}]{sabelfeld12}%
  \BibitemOpen
  \bibfield  {author} {\bibinfo {author} {\bibfnamefont {K.~K.}\ \bibnamefont
  {Sabelfeld}},\ }\href@noop {} {\emph {\bibinfo {title} {Random Fields and
  Stochastic Lagrangian Models. Analysis and Applications in Turbulence and
  Porous Media}}}\ (\bibinfo  {publisher} {Walter de Gruyter},\ \bibinfo
  {address} {Berlin},\ \bibinfo {year} {2012})\BibitemShut {NoStop}%
\bibitem [{\citenamefont {Walker}(1977)}]{walker77}%
  \BibitemOpen
  \bibfield  {author} {\bibinfo {author} {\bibfnamefont {A.~J.}\ \bibnamefont
  {Walker}},\ }\bibfield  {title} {\enquote {\bibinfo {title} {An efficient
  method for generating discrete random variables with general
  distributions},}\ }\href@noop {} {\bibfield  {journal} {\bibinfo  {journal}
  {ACM Trans. Math. Software}\ }\textbf {\bibinfo {volume} {3}},\ \bibinfo
  {pages} {253--256} (\bibinfo {year} {1977})}\BibitemShut {NoStop}%
\bibitem [{\citenamefont {Devroye}(1986)}]{devroye86}%
  \BibitemOpen
  \bibfield  {author} {\bibinfo {author} {\bibfnamefont {L.}~\bibnamefont
  {Devroye}},\ }\href@noop {} {\emph {\bibinfo {title} {Non-Uniform Random
  Variate Generation}}}\ (\bibinfo  {publisher} {Springer},\ \bibinfo {address}
  {N.Y.},\ \bibinfo {year} {1986})\BibitemShut {NoStop}%
\end{thebibliography}

%

\end{document}